\documentclass[a4paper,11pt]{article}
\pdfoutput=1

\usepackage{jcappub}

\usepackage[T1]{fontenc}

\usepackage[english]{babel}
\usepackage{amsfonts}
\usepackage{txfonts}
\usepackage{graphicx}
\usepackage{amssymb}
\usepackage{ae,aecompl}
\usepackage{fancyhdr}
\usepackage{multicol}
\usepackage{layout}
\usepackage{graphicx}
\usepackage{multirow}
\usepackage{times}
\usepackage{textcomp}
\usepackage{cprotect}
\usepackage{soul} 

\usepackage{comment}
\usepackage{color}

\usepackage[outdir=./figures/]{epstopdf}
\usepackage[export]{adjustbox}

\graphicspath{{./figures/}}

\newif\ifAMStwofonts
\AMStwofontstrue

\title{Tidal virialization of dark matter haloes with clustering dark energy}

\author[a,b,c,1]{Francesco Pace,\note{Corresponding author.}}
\author[d]{and Carlo Schimd}

\affiliation[a]{Dipartimento di Fisica ed Astronomia ``Augusto Righi'', Alma Mater Studiorum Universit\`a di Bologna, Via Gobetti 93/2, I-40129 Bologna, Italy}
\affiliation[b]{Dipartimento di Fisica, Universit\`a di Torino, Via P. Giuria 1, I-10125, Torino, Italy}
\affiliation[c]{Istituto Nazionale di Fisica Nucleare (INFN), Sezione di Torino, Via P. Giuria 1, I-10125, Torino, Italy}
\affiliation[d]{Aix Marseille Univ, CNRS, CNES, LAM, Marseille, France}

\emailAdd{francesco.pace9@unibo.it}
\emailAdd{carlo.schimd@lam.fr}

\date{\today}

\abstract{
We extend the analysis of Pace et al. \cite{Pace2019b} by considering the virialization process in the extended spherical collapse model for clustering dark-energy models, i.e., accounting for dark-energy fluctuations. Differently from the standard approach, here virialization is naturally achieved by properly modelling deviations from sphericity due to shear and rotation induced by tidal interactions. We investigate the time evolution of the virial overdensity $\Delta_\mathrm{vir}$ in seven clustering dynamical dark energy models and compare the results to the $\Lambda$CDM model and to the corresponding smooth dark-energy models.  
Taking into account all the appropriate corrections, we deduce the abundance of convergence peaks for Rubin Observatory-LSST and Euclid-like weak-lensing surveys, of Sunyaev-Zel'dovich peaks for a Simon Observatory-like CMB survey, and of X-ray peaks for an eROSITA-like survey. Despite the tiny differences in $\Delta_\mathrm{vir}$ between clustering and smooth dark-energy models, owing to the large volumes covered by these surveys, five out of seven clustering dark-energy models can be statistically distinguished from $\Lambda$CDM. The contribution of dark-energy fluctuation cannot be neglected, especially for the Chevallier-Polarski-Limber and Albrecht-Skordis models, provided the instrumental configurations provide high signal-to-noise ratio. These results are almost independent of the tidal virialization model.
}


\arxivnumber{2201.04193}

\begin{document}

\label{firstpage}

\maketitle
\flushbottom

\section{Introduction}\label{sect:intro}
The primary route towards the non-linear evolution of structures is the spherical collapse model \citep{Tomita1969,Gunn1972}. The formalism follows the evolution of an isolated, spherical patch of homogeneous overdensity, by firstly decoupling from the background expansion of the Universe to reach a maximum size (the turn-around) and subsequently to re-collapse to a singularity with null size and infinite density. This mathematical solution results from the approximation of exact spherical symmetry. Cosmic structures actually virialize and reach a finite size as a consequence of deviations from spherical symmetry and relaxation processes such as phase mixing, chaotic mixing, violent relaxation and Landau damping.

In the spherical collapse model, this is achieved by introducing the virial theorem. Virialization is completely understood in a purely matter-dominated Universe (Einstein-de Sitter model) and its study led to the well-known value of $\Delta_\mathrm{vir}=18\pi^2\simeq 178$ for the virial overdensity at collapse time and $18\pi^2(\tfrac{3}{4}+\frac{1}{2\pi})^2\simeq 147$ at virialization \cite{Lee2010,Lee2010c}, corresponding to the linearly-extrapolated overdensity $\delta_\mathrm{c}=3/20~(12\pi)^{2/3}\simeq 1.686$ at collapse \citep{Tomita1969} and $3/20~(6+9\pi)^{2/3} \simeq 1.58$  at virialization \cite{Lee2010,Lee2010c}, and a virial radius equal to half the turn-around radius. The exact values found in the Einstein-de Sitter model are time-independent.

Its simplest generalisation to the $\Lambda$CDM model originated a  debate \citep{Wang1998,Wang2006} because of how to model the virialization process in presence of a cosmological constant. 
The situation becomes even more complicated when extending the investigation to arbitrary dark energy models, which may take part in the virialization process \citep{Horellou2005,Battye2003,Iliev2001}. 
Dark energy indeed not only modifies the background expansion but it could also cluster, its perturbations directly affecting the evolution of matter perturbations. 
Virialization with clustering dark-energy has been investigated by \cite{Mota2004,Maor2005,Wang2006,Basse2011,Batista2017,Chang2018}, which used the virial theorem and considered the process if only dark matter or both dark matter and dark energy virialize. Though, the study by \cite{Maor2005} highlighted the difficulties and theoretical uncertainties of this description.

Since the problem is not settled yet, alternative views are welcome. A particularly interesting one is that proposed by \cite{Engineer2000} and lately refined by \cite{Shaw2008}. The idea is that departures from an exact spherical symmetry, such as shear and rotation induced by tidal interactions, will instead naturally yield to a virialization process.
This is done by considering an asymptotic expression for the shear and the rotation terms derived by the equations of motion of the system and adding corrections proportional to inverse powers of the overdensity. In \cite{Pace2019b} we have generalized the two original works developed for the EdS model to $\Lambda$CDM and dark-energy models with an arbitrary equation-of-state $w_\mathrm{de}$, proving that the results obtained with the new shear-and-rotation-based virialization model are qualitatively similar to those obtained with the standard recipe of \cite{Wang1998}: $\Delta_\mathrm{vir}$ is time-independent for the EdS model and becomes time-dependent for the other models, actually returning smaller values with \cite{Engineer2000} and \cite{Shaw2008} virialization recipes instead of the virial theorem. 
This has important consequences on the estimation of the abundance of dark-matter haloes, relevant for instance in lensing studies counting the number of convergence peaks.

In this work, we build upon the extended formalism outlined in \cite{Pace2019b} and study the effects of shear and rotation in clustering dark-energy models. Our work differs from previous works in the literature on the same subject 
\cite{Reischke2016a,Pace2017,Reischke2018} as in this case deviations from spherical symmetry are the cause of virialization, while before virialization was achieved by considering the virial theorem.

The values derived for the virial overdensity $\Delta_\mathrm{vir}$ will be used to study the impact on the number of convergence density peaks observed by Vera Rubin Observatory VRO-LSST\footnote{\href{https://www.lsst.org/}{https://www.lsst.org/}} and Euclid\footnote{\href{https://www.euclid-ec.org/}{https://www.euclid-ec.org/}}-like weak-lensing surveys, on the number of Sunyaev-Zeldovich peaks in a Simons Observatory\footnote{\href{https://simonsobservatory.org/}{https://simonsobservatory.org/}}-like survey and on X-ray peak counts in an eROSITA\footnote{\href{https://www.mpe.mpg.de/eROSITA}{https://www.mpe.mpg.de/eROSITA}}-like survey.
The influence of the clustering can be assessed by comparing the newly derived results with those of \cite{Pace2019b}. 
As at the moment $N$-body codes assume dark energy to just change the background expansion of the Universe, we can not compare our results directly with simulations. Therefore, as currently done in the literature, the reference model will be assumed to be $\Lambda$CDM and the importance of the deviations induced by the new physics and recipes will be assessed with respect to it.

The plan of this work is as follows: section~\ref{sect:SCM} reviews the formalism of the extended spherical collapse model and discusses its further generalisation to clustering dark-energy models, while section~\ref{sect:analysis} provides an extensive discussion of the properties of the virialization overdensity with respect to a smooth dark energy scenario and to previous works in the literature. These results, taking into account the appropriate modifications induced by the effects of clustering, will serve as the starting point of our analysis of the observable consequences of the set-up on weak lensing, X-ray and SZ surveys in section~\ref{sect:observables}. We finally conclude in section~\ref{sect:conclusions}.

As done in our previous work and to facilitate the comparison, we assume a spatially flat universe and a matter density 
parameter $\Omega_{\rm m,0}=0.3$.

\section{Virialization mechanism by shear and rotation}\label{sect:SCM}

In this section, we derive the equations used to study the evolution of the virial overdensity in the shear-rotation-induced formalism. We start in section~\ref{sect:SCM:fluid} by reviewing the equations of the extended spherical collapse model with the addition of shear and rotation, dubbed tidal virialization, for clustering dark-energy models. 
In section~\ref{sect:SCM:s2o2}, we discuss how to modify the parameterization of the shear and rotation terms following the approach of \cite{Engineer2000}.  
We will not discuss the formulation of \cite{Shaw2008} since, as shown in \cite{Pace2019b}, it produces a very small $\Delta_\mathrm{vir}$ which leads to convergence peak counts at odds with $\Lambda$CDM simulations and observations.

\subsection{The fluid approach}\label{sect:SCM:fluid}

The discussion of this section is largely based on \cite{Pace2017a} where a detailed numerical implementation has been outlined. Here, we limit ourselves to present the equations as they are implemented in practice (as a set of first-order ordinary differential equations), but we will combine the equations for matter perturbations into a single second-order equation which will be useful for the following discussion.\footnote{For clarity of exposition, we write the equations in terms of the dimensionless density perturbation $\delta$. For numerical stability, it is, though, better to solve them for $1/\delta$ and this is the actual implementation of the modified equations.}

A simple and straightforward way to derive the equations of motion of matter and dark energy in a general relativistic framework is the fluid approach. Both matter and dark energy perturbations are described in terms of the density $\delta\rho$, pressure $\delta P$ and velocity perturbations $\boldsymbol{u}$.\footnote{By matter we consider only baryons and the cold dark matter component, as we consider structure formation at late times, where the effects of dark energy and its perturbations are more important.} From the continuity and Euler equations one derives first-order equations for the density and velocity perturbation, respectively. In general, these equations are complicated partial differential equations in both time and space, but a standard assumption is that of assuming a top-hat profile: the density profile of the perturbation is spatially constant and it can evolve only in time. From a physically point of view, this is equivalent to consider a window function with a size larger than that of the perturbation. This simplification, albeit crude, leads to results that are in agreement with $N$-body simulations \cite{Pace2010,Tarrant2012,Pace2014} and to analytical results, as for the EdS model \cite{Kihara1968}, which allow to test the reliability of the code \cite{Pace2017a}.

The equations we will consider are the following \citep{Abramo2007,Abramo2008,Abramo2009a,Abramo2009b,Pace2017a}
\begin{subequations}
 \begin{align}
  \delta_\mathrm{m}^{\prime} + (1+\delta_\mathrm{m})\tilde{\theta} = &\, 0\,, \label{eqn:deltam} \\
  \tilde{\theta}^{\prime} + \left(2+\frac{H^{\prime}}{H}\right)\tilde{\theta} + 
  \frac{\tilde{\theta}^2}{3} + 
  \frac{3}{2}\left[\Omega_\mathrm{m}(a)\delta_\mathrm{m}+(1+3s_{\rm eff})\Omega_\mathrm{de}(a)\delta_\mathrm{de}\right] + 
  \tilde{\sigma}^2-\tilde{\omega}^2 = &\, 0\,, \label{eqn:theta} \\
  \delta_\mathrm{de}^{\prime} + 3(s_{\rm eff}-w_\mathrm{de})\delta_\mathrm{de} + 
  \left[1+w_\mathrm{de}+\left(1+s_{\rm eff}\right)\delta_\mathrm{de}\right]\tilde{\theta} = &\, 0\,, 
  \label{eqn:deltade}
 \end{align}
\end{subequations}
where a prime represents the derivative with respect to $\ln{a}$, $H$ is the Hubble function, $\tilde{\theta}=\theta/H$, $w_\mathrm{de}$ the background equation-of-state and $c_{\rm eff}^2=c^2s_{\rm eff}=\delta P_\mathrm{de}/\delta\rho_\mathrm{de}$ the effective speed-of-sound of perturbations with $c$ the speed of light. For a quintessence model, $s_{\rm eff}=1$ and perturbations are strongly suppressed and negligible: this is equivalent to consider dark energy only affecting the background and justifies why $N$-body simulations only consider this case. We will, therefore, assume a more general model where $s_{\rm eff}\ll 1$ and dark energy perturbations become relatively important for the matter dynamics, albeit usually always subdominant 
\citep{Batista2013}. 
A small speed-of-sound can be achieved in models with two scalar fields \citep{Lim2010,Sebastiani2017,Cognola2016} and using the effective field theory \citep{Creminelli2009}. Matter and dark energy density parameters are indicated with $\Omega_\mathrm{m}(a)$ and $\Omega_\mathrm{de}(a)$, respectively.

We have defined $\sigma^2\equiv\sigma_{ij}\sigma^{ij}$ and $\omega^2\equiv\omega_{ij}\omega^{ij}$ as the squared amplitude of the shear and rotation tensors, respectively, and they represent the symmetric traceless and anti-symmetric component of the derivative of the peculiar velocity $\boldsymbol{u}$,
\begin{equation}\label{eqn:veltensor}
 \partial_i u_j = \frac{1}{3}\theta\delta_{ij}+\sigma_{ij}+\omega_{ij}\,,
\end{equation}
and $\theta\equiv\vec{\nabla}\cdot\boldsymbol{u}$ accounts for the isotropic expansion. The shear and rotation tensors read
\begin{equation}
 \sigma_{ij} = \frac{1}{2}\left(\partial_i u_j + \partial_j u_i\right) - \frac{1}{3}\theta\delta_{ij}\,, \quad 
 \omega_{ij} = \frac{1}{2}\left(\partial_i u_j - \partial_j u_i\right)\,.
\end{equation}

In the previous equations, we have assumed that the term $\tilde{\sigma}^2-\tilde{\omega}^2=(\sigma^2-\omega^2)/H^2$ 
affects both matter and dark energy perturbations, leading to $\tilde{\theta}_\mathrm{m}=\tilde{\theta}_\mathrm{de}=\tilde{\theta}$. As shown in \citep{Pace2014b,Pace2017}, when the shear and rotation terms are considered only for matter, this setup does not yield appreciable differences in the evolution of perturbations. 
This is true for models in which deviations from sphericity are proportional to $\delta$ \cite{DelPopolo2013a,DelPopolo2013b,DelPopolo2013c} and models in which a more realistic description based on the Zel'dovich approximation is assumed \citep{Reischke2018}. However, differently from the modelling above, the formulation of \cite{Engineer2000} and \cite{Shaw2008} does not introduce any mass dependence making impossible a direct comparison with previous recipes. Furthermore, here the additional term $\tilde{\sigma}^2-\tilde{\omega}^2$ is meant directly support virialization, which was not the case for \cite{DelPopolo2013a,DelPopolo2013b,DelPopolo2013c,Reischke2018}, where stable structures were achieved by considering the virial theorem. For this reason, we find it appropriate and physically realistic to assume that also dark energy perturbations are directly affected by the virialization term.

Eqs.~(\ref{eqn:deltam}) and (\ref{eqn:theta}) can be combined to obtain the well-known non-linear equation for matter perturbations \citep{Pace2010,Pace2017a}, which is now augmented by dark energy perturbations which arise in the Poisson equation. The resulting equation is
\begin{equation}\label{eqn:nldelta}
 \delta_\mathrm{m}^{\prime\prime} + \left(2+\frac{H^{\prime}}{H}\right)\delta_\mathrm{m}^{\prime} - 
 \frac{4}{3}\frac{{\delta_\mathrm{m}^{\prime}}^2}{1+\delta_\mathrm{m}} - \frac{3}{2}(1+\delta_\mathrm{m})
 \left[\Omega_\mathrm{m}(a)\delta_\mathrm{m}+(1+3s_{\rm eff})\Omega_\mathrm{de}(a)\delta_\mathrm{de}\right] = 
 (1+\delta_\mathrm{m})(\tilde{\sigma}^2-\tilde{\omega}^2)\,.
\end{equation}

One can, in principle, apply the same argument to the dark energy component, however, this makes the final equation rather cumbersome, as it contains derivatives of the equation-of-state $w_\mathrm{de}$ and the effective speed-of-sound $s_{\rm eff}$. Therefore, we do not derive this equation since matter perturbations only depend on $\delta_\mathrm{de}$.

\subsection{Tidal virialization in presence of dark-energy fluctuations}\label{sect:SCM:s2o2}

We focus here on the virialization process induced by the shear and rotation term \cite{Engineer2000,Pace2019b}. 
As in these two studies, we use the shorthand notation $S=\tilde{\sigma}^2-\tilde{\omega}^2$ to indicate the shear-rotation-induced virialization term.

When virialization takes place, the system is in equilibrium, the overdensity $\delta_\mathrm{m}$ tends to a constant value and $\delta_\mathrm{m}^{\prime}\approx 0$, so that Eq.~(\ref{eqn:nldelta}) reduces to
\begin{equation}\label{eqn:S2}
 S_2 \approx -\frac{3}{2}\Omega_\mathrm{m}\delta_\mathrm{m}\left[
              1+\frac{\Omega_\mathrm{de}(a)}{\Omega_\mathrm{m}(a)}\frac{\delta_\mathrm{de}}{\delta_\mathrm{m}}(1+3s_{\rm eff})
              \right]\,,
\end{equation}
where the second term in parenthesis represents the corrections introduced by the perturbations of the dark energy component with respect to a smooth dark energy component.\footnote{We assume that when matter virialises the same holds for the dark energy component, $\delta_\mathrm{de}\approx$ constant. Of course this might not be the case, as dark energy perturbations could continue to evolve and not virialize at all. Since in general these details are unknown, it seems reasonable to halt the evolution of $\delta_\mathrm{de}$ at virialization.} Using the same notation as \cite{Pace2019b}, we refer to this as model $S_2$.

Eq.~(\ref{eqn:nldelta}) can also be recast in terms of the (effective) radius of the perturbation $R$. 
Following the discussion of \cite{Batista2013}, we define
\begin{equation}
 M_\mathrm{m} = \frac{4\pi}{3}\bar{\rho}_\mathrm{m}(a)\left(1+\delta_\mathrm{m}\right)R^3\,, \quad 
 M_\mathrm{de} = \frac{4\pi}{3}\bar{\rho}_\mathrm{de}(a)\delta_\mathrm{de}R^3\,, \quad 
 M_{\rm bck,de} = \frac{4\pi}{3}\bar{\rho}_\mathrm{de}(a)R^3\,,
\end{equation}
where $M_\mathrm{m}$ and $M_\mathrm{de}$ are the matter and the dark energy component of the perturbation, such that $M_{\rm tot} = M_\mathrm{m} + M_\mathrm{de}(1+3s_{\rm eff})$ and $M_{\rm bck,de}$ is the dark energy contribution to the background and $\bar{\rho}_\mathrm{m}(a)$ and $\bar{\rho}_\mathrm{de}(a)$ are the background density for matter and dark energy, respectively. The equation of motion for the radius is
\begin{equation}\label{eqn:Rddot}
 \ddot{R} = -\frac{GM_\mathrm{m}}{R^2} - \frac{GM_\mathrm{de}}{R^2}(1+3s_{\rm eff}) - 
             \frac{GM_{\rm bck,de}}{R^2}(1+3w_\mathrm{de}) - \frac{R}{3}H^2S\,,
\end{equation}
which generalises Eq.~(2.5) of \cite{Pace2019b} and Eq.~(2.4) of \cite{Batista2017} as it considers arbitrary effective speed-of-sounds $s_{\rm eff}$. In Eq.~(\ref{eqn:Rddot}), the relation between the matter overdensity $\delta_\mathrm{m}$ and the effective radius $R$ is given by
\begin{equation}
 1 + \delta_\mathrm{m} = \frac{2GM_\mathrm{m}a^3}{\Omega_{\rm m,0}H_0^2R^3} = \lambda\frac{a^3}{R^3}\,.
\end{equation}

It is first important to understand the limits of this equation in special cases.
When dark energy is smooth, $\delta_\mathrm{de}=0$ so that $M_\mathrm{de}$ vanishes. In this case, Eq.~(\ref{eqn:Rddot}) reduces to Eq.~(2.5) of \cite{Pace2019b}. When $s_{\rm eff}=0$, we recover Eq.~(2.4) of \cite{Batista2017} and for the case $s_{\rm eff}=w_{\rm eff}$, one can redefine the mass associated to dark energy perturbations as $M_\mathrm{de}=\tfrac{4\pi}{3}\rho_\mathrm{de}(a)R^3$, where $\rho_\mathrm{de}(a)=\bar{\rho}_\mathrm{de}(a)(1+\delta_\mathrm{de})$.

Eq.\eqref{eqn:Rddot} further allows a discussion similar to \cite{Batista2013} on how are dark energy perturbations considered. In \cite{Batista2013}, the authors discussed the contribution of the dark energy mass to the halo mass function, according to whether dark energy virializes and, if it happens, on which time-scale. This contribution boils down to how the total mass of the halo is defined, which ultimately depends on the definition of the dark-energy mass. 
While the dark-matter mass is defined considering the contribution of the background ($\bar{\rho}_\mathrm{m}$) and of the perturbations 
($\delta\rho_\mathrm{m}$), for dark-energy there are two options: either one can consider only the contribution of the perturbations ($\delta\rho_\mathrm{de}$), or one can also include the background ($\bar{\rho}_\mathrm{de}$) as for the dark matter case. 
Which definition is better or more appropriate is rather arbitrary, however, it seems reasonable to consider only dark-energy perturbations contributing to the total halo mass, as normally the latter does not include the background contribution of dark energy.\footnote{This is the case, for example, of $N$-body simulations where the cosmological constant does not enter in the definition of the halo mass.} The situation becomes even more uncertain when the dark energy component does not fully cluster.

This means that while $M_\mathrm{m}$ is conserved, nor $M_\mathrm{de}$ and $M_{\rm tot}$ are. In this case, the virialization time is defined when the moment of inertia of a sphere $I=2M_{\rm tot}R^2/5$ of non-relativistic particles is in a steady state, i.e. $\mathrm{d}^2I/\mathrm{d}t^2=0$, yielding \cite{Basse2012,Batista2017}
\begin{equation}
 \frac{1}{2M_{\rm tot}}\frac{\mathrm{d}^2M_{\rm tot}}{\mathrm{d}t^2} +
 \frac{2}{M_{\rm tot}R}\frac{\mathrm{d}M_{\rm tot}}{\mathrm{d}t}\frac{\mathrm{d}R}{\mathrm{d}t} + 
 \frac{1}{R^2}\left(\frac{\mathrm{d}R}{\mathrm{d}t}\right)^2 + 
 \frac{1}{R}\frac{\mathrm{d}^2R}{\mathrm{d}t^2} = 0\,.
\end{equation}

After the previous digression, we discuss again Eq.~(\ref{eqn:Rddot}). When virialization is reached, $R\rightarrow R_\mathrm{vir}$ where $R_\mathrm{vir}$ is the virialization radius and $\dot{R}=\ddot{R}\approx 0$, leading to
\begin{equation}\label{eqn:S1}
 \begin{split}
  S_1 \approx & -\frac{3GM_\mathrm{m}}{H^2R^3}\left[1 + \frac{M_\mathrm{de}}{M_\mathrm{m}}(1+3s_{\rm eff}) + 
  \frac{M_{\rm bck,de}}{M_\mathrm{m}}(1+3w_\mathrm{de})\right]\,,\\
  \approx & -\frac{3}{2}\Omega_\mathrm{m}(a)(1+\delta_\mathrm{m})
  \left[1+\frac{\Omega_\mathrm{de}(a)}{\Omega_\mathrm{m}(a)}
  \frac{\delta_\mathrm{de}}{1+\delta_\mathrm{m}}(1+3s_{\rm eff})\right]\,,
 \end{split}
\end{equation}
where we neglected the contribution of the background dark energy component, as negligible with respect to the other two terms. The term in the parenthesis quantifies the contribution of dark energy perturbations, which, for a late time virialization, are usually of the order of a few percent \citep{Batista2013,Batista2017}. We call this model $S_1$. As shown in \cite{Pace2019b}, expressions $S_1$ and $S_2$ will be the same when $\delta\gg 1$, but differ for $\delta\lesssim 1$.

Following \cite{Engineer2000}, we consider the virial term as a function of the overdensity only and expand it up to the second order in inverse powers of the overdensity. While for smooth dark energy models it is obvious that $\delta=\delta_\mathrm{m}$, it is less clear how to define it in clustering dark-energy models. From the discussion above, it appears that the expressions from \cite{Pace2019b} receive a contribution of the order $M_\mathrm{de}/M_\mathrm{m}$, which is, in general, up to a few percent at late times when dark energy dominates and totally negligible before \citep{Batista2013,Batista2017}. 
Considering or ignoring it, would, therefore, not change our conclusions, but only, possibly, the exact value of the virial overdensity $\Delta_\mathrm{vir}$. We will, therefore, expand the shear-rotation-induced virialization term in inverse powers of
\begin{equation}\label{eqn:delta}
 \delta \equiv \delta_\mathrm{m} + \frac{\Omega_\mathrm{de}(a)}{\Omega_\mathrm{m}(a)}\left(1+3s_{\rm eff}\right)\delta_\mathrm{de}\,.
\end{equation}
Note that this definition reduces to $\delta_\mathrm{m}$ when $\delta_\mathrm{de}=0$ and this shows that it is easy and straightforward to go back to the case of smooth dark energy without invoking additional properties of the dark energy fluid involved in the virialization process.

The virialization term following both procedures then reads
\begin{align}
 S_1 = &\, -\frac{3}{2}\Omega_\mathrm{m}(a)(1+\delta) - \frac{A}{\delta} + \frac{B}{\delta^2}\,,\\
 S_2 = &\, -\frac{3}{2}\Omega_\mathrm{m}(a)\delta - \frac{A}{\delta} + \frac{B}{\delta^2}\,,
\end{align}
where $A$ and $B$ are arbitrary constants.

Following the same procedure of \cite{Engineer2000,Pace2019b}, we can, therefore, write the equation of motion for matter perturbations in the presence of a virialization term $S$ induced by shear and rotation as
\begin{subequations}\label{eqn:deltaS}
 \begin{align}
  & \delta_\mathrm{m}^{\prime\prime} + \left(2+\frac{H^{\prime}}{H}\right)\delta_\mathrm{m}^{\prime} - 
    \frac{4}{3}\frac{{\delta_\mathrm{m}^{\prime}}^2}{1+\delta_\mathrm{m}} + 
    \frac{3}{2}\Omega_\mathrm{m}(a)(1+\delta_\mathrm{m}) = 
    (1+\delta_\mathrm{m})\left(-\frac{1}{\delta}+\frac{q}{\delta^2}\right)\,,\\
  & \delta_\mathrm{m}^{\prime\prime} + \left(2+\frac{H^{\prime}}{H}\right)\delta_\mathrm{m}^{\prime} - 
    \frac{4}{3}\frac{{\delta_\mathrm{m}^{\prime}}^2}{1+\delta_\mathrm{m}} = 
    (1+\delta_\mathrm{m})\left(-\frac{1}{\delta}+\frac{q}{\delta^2}\right)\,,
 \end{align}
\end{subequations}
where $q=B/A^2$ is a free parameter of the model. These equations are formally identical to Eqs.~(2.11) of \cite{Pace2019b} and we can see that, except for small modifications due to the contribution of dark energy perturbations, the physics of virialization is the same as in \cite{Pace2019b}. The whole dynamics is, therefore, dictated by the parameter $q$ only and different values will lead to a different evolution of matter perturbations and hence to a different virial overdensity $\Delta_\mathrm{vir}$. The exact dynamics of the two expressions will differ from each other only at intermediate values of $\delta_\mathrm{m}$ and this will lead to small, but appreciable differences in $\Delta_\mathrm{vir}$, as already discussed in \cite{Pace2019b}.

Since the final equations (\ref{eqn:deltaS}) resemble very closely the case of smooth dark energy, it may seem that they are a trivial result and give us no new understanding of the physics involved. In reality, this is an important result, as the whole derivation is useful in understanding the virialization process in clustering dark-energy models, being physically neat and not requiring additional hypotheses about whether dark energy virialises or not with the dark matter component, as in \cite{Maor2005,Wang2006}. In addition, a value of $s_{\rm eff}\neq 0$ does not lead to particular complications in our derivation. This is different and somewhat simpler, for example, than the discussion in \cite{Maor2005}, where the authors showed that the final result depends on the virialization properties of dark energy. We will make a detailed comparison of the different recipes for virialization in clustering and smooth dark energy models in a following work.

Finally, we can derive the equation of motion for the evolution of the radius $R$ in the limit $\delta\gg 1$. As in 
\cite{Pace2019b,Engineer2000}, it is better to rewrite it in terms of the dimensionless radius $y$, where $R=yr_\mathrm{vir}$, with $r_\mathrm{vir}$ representing the virial radius. Following the same steps of \cite{Pace2019b} and using the same definitions and notation, we can write the equation for $y$, which reads
\begin{subequations}\label{eqn:yq}
 \begin{align}
  y^{\prime\prime}+\frac{H^{\prime}}{H}y^{\prime} = &\, -\frac{1}{2}\Omega_\mathrm{de}(a)(1+3w_\mathrm{de})y + \frac{y}{3}
   \left[\frac{\Omega_{\rm m,0}}{2a^3\tilde{\epsilon}}y^3
  -q\frac{\Omega_{\rm m,0}^2}{4a^6\tilde{\epsilon}^2}y^6\right]\,,\\
  y^{\prime\prime}+\frac{H^{\prime}}{H}y^{\prime} = &\, 
  -\frac{1}{2}\left[\Omega_\mathrm{m}(a)+\Omega_\mathrm{de}(a)(1+3w_\mathrm{de})\right]y + 
  \frac{y}{3}\left[\frac{\Omega_{\rm m,0}}{2a^3\tilde{\epsilon}}y^3-
  q\frac{\Omega_{\rm m,0}^2}{4a^6\tilde{\epsilon}^2}y^6\right]\,,
 \end{align}
\end{subequations}
where we defined $\tilde{\epsilon}\equiv[1+(1+3s_{\rm eff})\epsilon]$ with $\epsilon\equiv M_\mathrm{de}/M_\mathrm{m}$. 
Differently from \cite{Batista2013,Batista2017}, we define $\epsilon$ as a generic function of time, rather than evaluated at virialization only.

One can immediately see that, formally, the evolution of the radius is similar to what found in \cite{Pace2019b}, with the additional contribution of dark energy perturbations only in the virialization term due to the small correction parameterised by the function $\epsilon$. Since Eqs.~(\ref{eqn:yq}) are solved backwards with initial conditions $y=1$ at collapse redshift $z_\mathrm{c}$, we expect higher differences with respect to the smooth case only for a late time collapse (in agreement with the evolution of the overdensity $\delta_\mathrm{m}$). Modifications are nevertheless small and results obtained in our previous work still hold qualitatively, albeit the exact numerical results will be slightly modified by the inclusion of dark energy perturbations.

The whole derivation of the previous equations is based on the underlying assumption that $S=S(\delta)$ only. This assumption has been justified by \cite{Engineer2000} by assuming the validity of the stable clustering Ansatz in the highly non-linear regime. While this is a simple approximation, it has been verified for the EdS and the $\Lambda$CDM models \citep{Padmanabhan1996b,Munshi1997}. This Ansatz is valid also in this work, as we are interested in regimes where $\delta\gg 1$. One of the consequences of the stable clustering Ansatz is that in the non-linear regime the rescaled pairwise velocity $h\rightarrow 1$. Kanekar \cite{Kanekar2000} showed that stable clustering is viable, but for $\delta\rightarrow\infty$, $0\leq h\leq 1/2$. This does not affect our results, as explained in \cite{Pace2019b}, to which we refer for a detailed discussion. Further to this, \cite{Batista2013} showed that clustering dark-energy models are closer to the $\Lambda$CDM solution than smooth models. This assures us that we can safely use stable clustering also in this work, as the free parameter $q$ is inferred at the peak of the rescaled pairwise velocity.

\section{Analysis}\label{sect:analysis}

In this section we present the results for the virial overdensity $\Delta_\mathrm{vir}$ for the same dark energy models studied in our previous work \cite{Pace2019b},\footnote{They are referred as DE1 and DE2 for models with constant equation-of-state respectively larger and smaller than $-1$, CPL for Chevallier-Polarski-Linder \cite{Chevallier2001,Linder2003}, ODE \cite{Pace2012,Pan2017,Panotopoulos2018}, CNR \citep{Copeland2000}, 2EXP \citep{Barreiro2000}, and AS for Albrecht-Skordis \cite{Albrecht2000}. \label{ftn:models}} and compare our results with those found using the approach of \cite{Engineer2000}. Since the contribution of dark energy is small and limited to late times, where it is the dominant component in the cosmic energy budget, we will not repeat the full analysis of \cite{Pace2019b}. We will, therefore, not show the detailed evolution of the virialization term $S(\delta)$ and of the dimensionless radius $y$. 
That's because we will not see qualitative differences with previous results. We will instead focus on the quantities more relevant for the following discussion and for comparison with other works in the literature.

In section~\ref{sect:epsilon} we shall first discuss the evolution of $\epsilon \equiv M_\mathrm{de}/M_\mathrm{m}$ to quantitatively asses dark energy corrections to matter perturbations. This discussion will help us to present the evolution of the perturbed equation-of-state $w_\mathrm{c}$ in section~\ref{sect:eos}, which we can compare with the corresponding background evolution. We finally move on to discuss the properties of the virial overdensity $\Delta_\mathrm{vir}$, which represents one of the main results of this work, in section~\ref{sect:DeltaV}.

Since we need to fix the value of the free parameter $q$, for simplicity we will assume the same values used in \cite{Pace2019b} (see their table~2).

\subsection{Virial radius with dark energy fluctuations}\label{sect:epsilon}

Here, we discuss the effects of dark energy perturbations on the virial radius $R_\mathrm{vir}$, presenting the evolution of the parameter $\epsilon=M_\mathrm{de}/M_\mathrm{m}$ defined in section~\ref{sect:SCM:s2o2}. Ultimately, this determines how much $\Delta_\mathrm{vir}$ changes from the case of smooth dark energy.

As explained before, we assume that both matter and dark energy virialize at the same time. Therefore, we need to solve Eqs.~(\ref{eqn:deltaS}) with the equation for the dark energy perturbations (\ref{eqn:deltade}) and a modified equation for peculiar velocities, once we made explicit the virialization term $S$. This equation reads
\begin{align}
 & \tilde{\theta}^{\prime} + \left(2+\frac{H^{\prime}}{H}\right)\tilde{\theta} + \frac{\tilde{\theta}^2}{3} - 
   \frac{3}{2}\Omega_\mathrm{m}(a) = \frac{1}{\delta} - \frac{q}{\delta^2}\,, \\
 & \tilde{\theta}^{\prime} + \left(2+\frac{H^{\prime}}{H}\right)\tilde{\theta} + \frac{\tilde{\theta}^2}{3} = 
   \frac{1}{\delta} - \frac{q}{\delta^2} \,,
\end{align}
for $S_1$ and $S_2$, respectively.

\begin{figure}[t]
 \centering
 \includegraphics[scale=0.6]{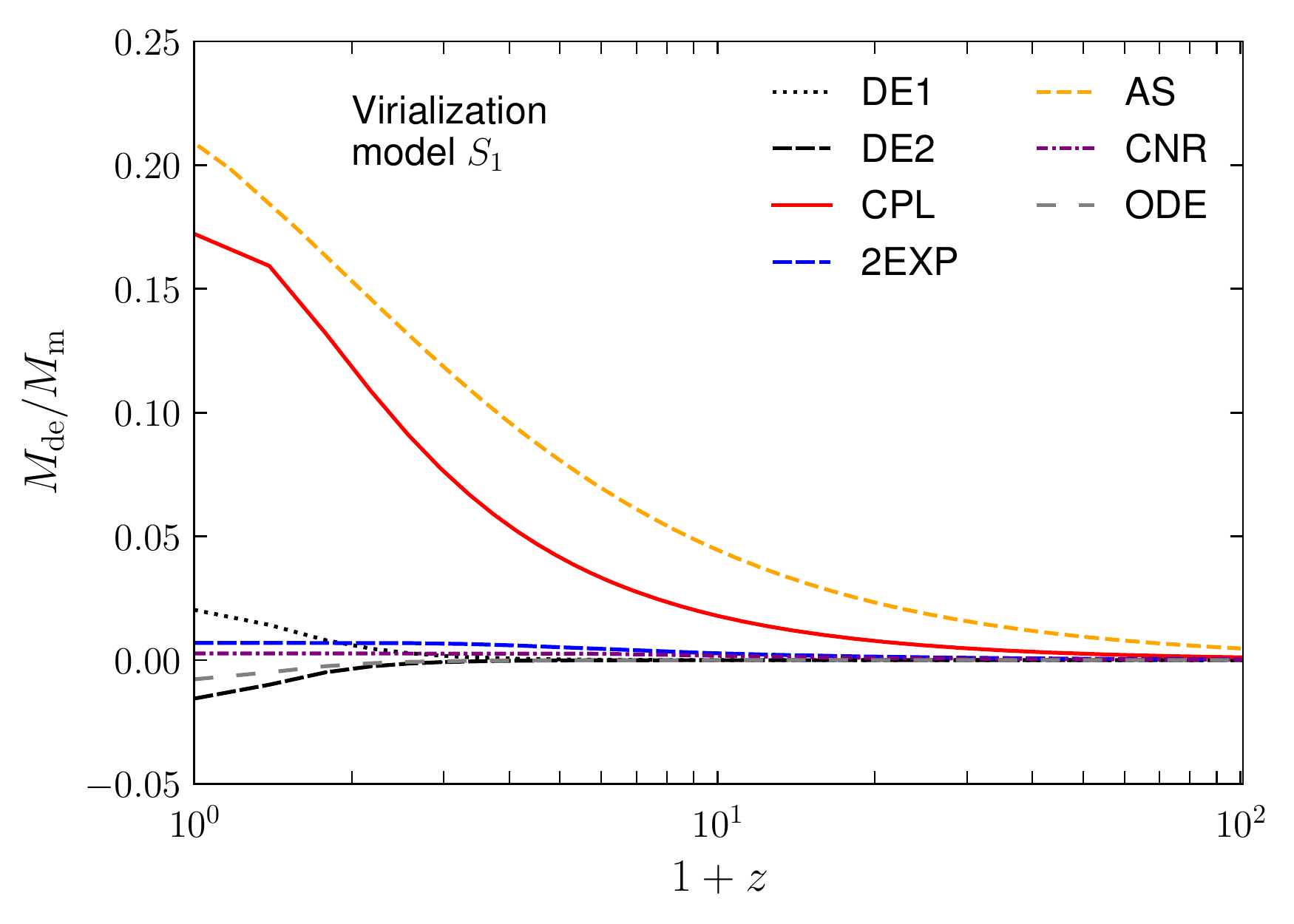}
 \caption{Time evolution of $\epsilon = M_\mathrm{de}/M_\mathrm{m}$ for the virialization model $S_1$ as a function of redshift $z$ for the dark energy models for a collapse occurring at $z=0$. Almost identical behaviour is found for the virialization model $S_2$.}
 \label{fig:epsilon}
\end{figure}

In figure~\ref{fig:epsilon} we present the behaviour of $\epsilon$ for a collapse at $z=0$ for the recipe $S_1$. The result for recipe $S_2$ is almost indistinguishable, hence not shown. This means that the ratio $\delta_\mathrm{de}/(1+\delta_\mathrm{m})$ is largely unaffected by the specific virialization recipe, as both matter and dark energy are equally affected by virialization. In general, $\epsilon>0$ (hence $\delta_\mathrm{de}>0$) except for the models DE2 and ODE, which are in the phantom regime. Why this is the case, it is easy to see considering the relation between $\delta_\mathrm{de}$ and $\delta_\mathrm{m}$ in the matter dominated regime (and assuming, as we do here, $s_{\rm eff}=0$) \citep{Abramo2008,Batista2013}
\begin{equation}
 \delta_\mathrm{de} = \frac{1+w_\mathrm{de}}{1-3w_\mathrm{de}}\delta_\mathrm{m}\,.
\end{equation}

This means that dark energy perturbations grow proportionally to dark matter perturbations. This is true for models where $w_\mathrm{de}>-1$, while for phantom models, $\delta_\mathrm{de}$ is negative and decreases. In other words, for quintessence-like models, matter overdensities are associated to dark energy overdensities, while in the phantom case they are associated to dark energy voids. Therefore, for models DE2 and ODE, $\delta_{\rm de, ini}<0$ and so is $\epsilon$.

For most models, dark energy perturbations are relatively small, about a few percent. Noticeable exceptions are the CPL (red solid line) and AS (orange dashed line) models. Relatively to matter, dark energy perturbations are around 20\% which, as we will see in the next sections, will lead to important modifications to the quantities we are going to examine.

To explain the behaviour of these two models, we need to consider the background equation-of-state $w_\mathrm{de}$, shown in figure~5 in \cite{Pace2019b}. Excluding models DE1 and DE2 having a constant $w_\mathrm{de}\neq -1$, CPL and AS show a very rapid departure from $w=-1$ at late times leading to important modifications at the background level which reflect in Eq.~(\ref{eqn:deltade}). The other models depart from $w=-1$ at later times and have a rapid transition to $w\approx 0$. CPL and AS instead, have a gradual transition which spans a relatively large range in redshift and this reflects on the evolution of perturbations. From another point of view, having for longer time a lower rate of expansion allows perturbations to grow more easily with respect to other models where growth is hampered by the accelerated expansion.

\subsection{Perturbed equation-of-state}\label{sect:eos}
In the previous section we showed that the difference between the two models of virialization is usually not important for the evolution of dark energy perturbations. This is an indication of the fact that dark energy perturbations are relatively independent of the exact equation used for $\theta$. In this section we consider how dark energy perturbations affect the equation of state by studying the evolution of the perturbed equation-of-state of clustering dark energy cosmologies.

The perturbed equation-of-state is defined as
\begin{equation}\label{eqn:wc}
 w_\mathrm{c} = \frac{1}{c^2}\frac{P_\mathrm{de}+\delta P_\mathrm{de}}{\rho_\mathrm{de}+\delta_\mathrm{de}} 
              = w_\mathrm{de} + (s_{\rm eff}-w_\mathrm{de}) \frac{\delta_\mathrm{de}}{1+\delta_\mathrm{de}}\,,
\end{equation}
which ranges between $w_\mathrm{de}$ (when perturbations are negligible, early times) and $s_{\rm eff}$ (when perturbations dominate, late times). For quintessence models, $s_{\rm eff}=1$ and dark energy perturbations are very small and important only on super-horizon scales. In the regime of interest to this work, this implies $w_\mathrm{c}\approx w_\mathrm{de}$, i.e., smooth models. In the following, therefore, we will consider more generic models with $s_{\rm eff}\approx 0$, which can be achieved for $k$-essence cosmologies and lead to stronger effects as dark energy will behave similarly to dark matter and fully cluster.

In this specific case, the perturbed equation-of-state (\ref{eqn:wc}) simplifies to
\begin{equation}
 w_\mathrm{c} = w_\mathrm{de}\left(1 - \frac{\delta_\mathrm{de}}{1+\delta_\mathrm{de}}\right) = 
 \frac{w_\mathrm{de}}{1+\delta_\mathrm{de}}\,.
\end{equation}
In this particular setup, the behaviour of $w_\mathrm{c}$ is directly dictated by dark energy perturbations, scaled by the value of the background equation-of-state $w_\mathrm{de}$.

We will show the evolution of the perturbed equation-of-state $w_\mathrm{c}$ in two different cases: the first for the case without virialization, i.e., the system collapses completely to a point and density perturbations diverge, and the second by taking into account the virialization term, where both dark matter and dark energy perturbations reach a large but finite value.

\begin{figure}[!t]
 \centering
 \includegraphics[scale=0.42]{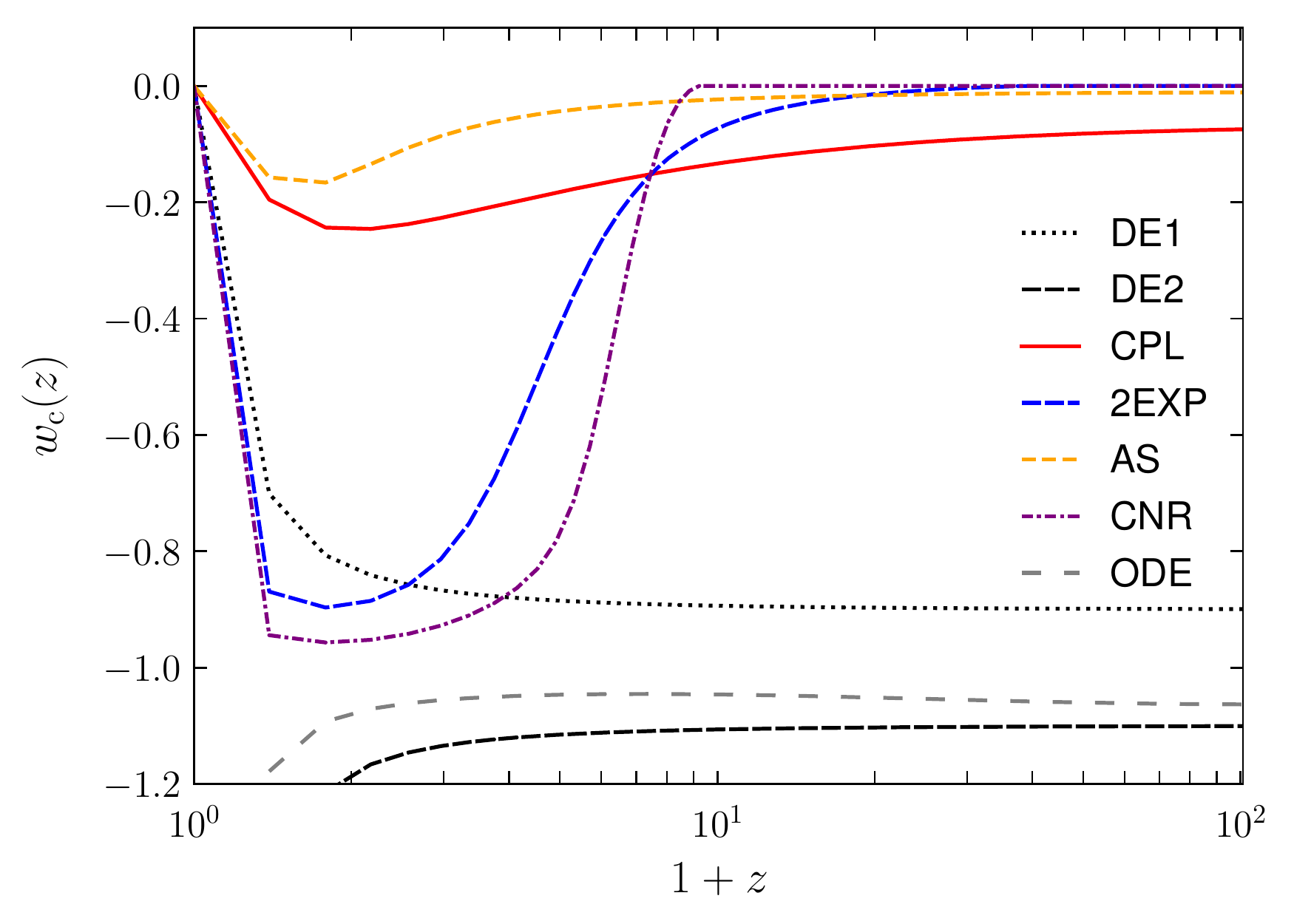}
 \includegraphics[scale=0.42]{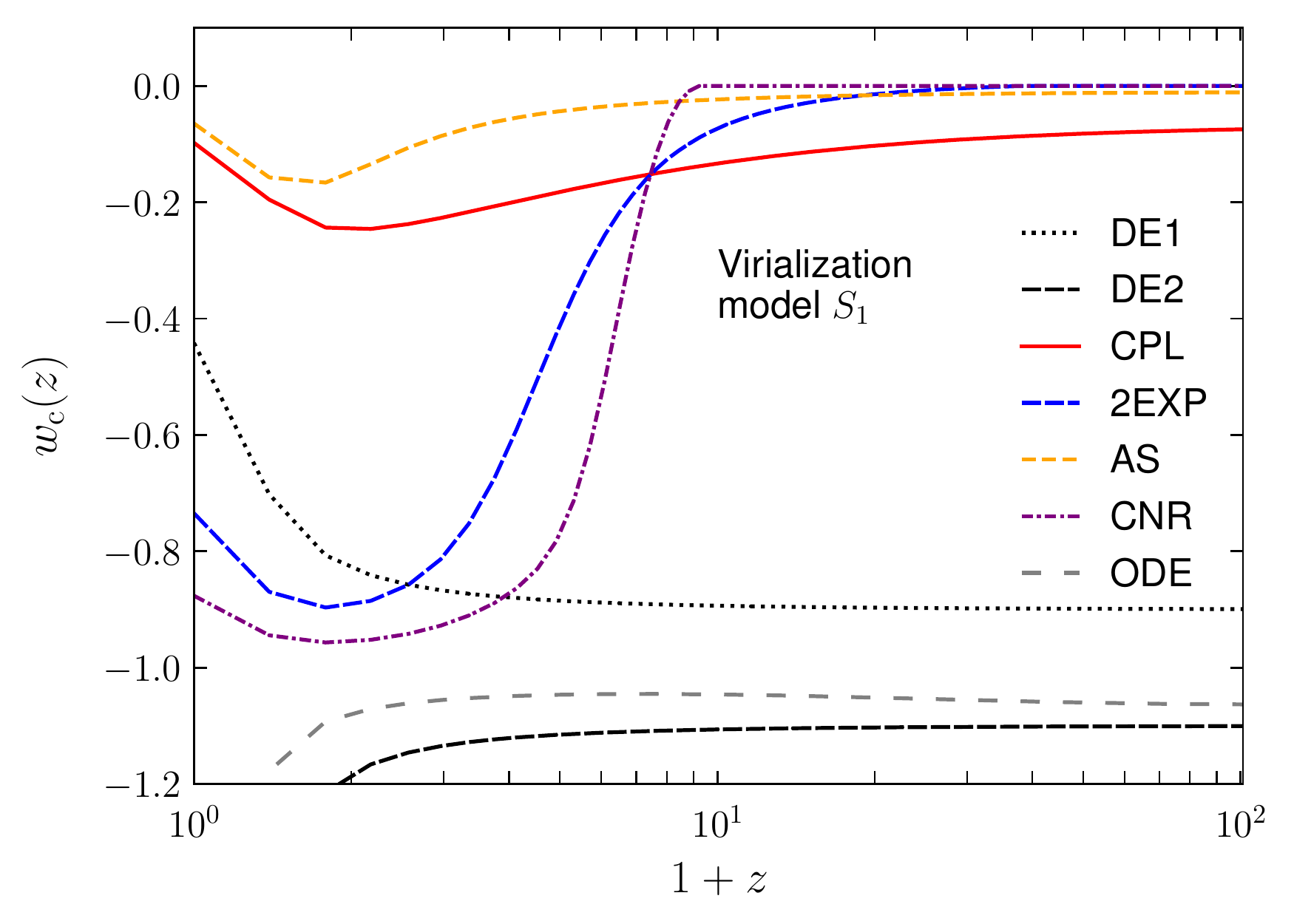}
 \caption{Time evolution of the perturbed equation-of-state $w_\mathrm{c}(z)$, Equation~(\ref{eqn:wc}), accounting or not for virialization (right and left panels) according to the $S_1$ virialization recipe induced by the shear-rotation term, for the different dark energy models (see legend; colour/line style as in figure~\ref{fig:epsilon}). We assumed that the system collapses at $z=0$.}
 \label{fig:wc}
\end{figure}

Our results for a collapsing system at $z=0$ are presented in figure~\ref{fig:wc}, where we show the evolution of $w_{\rm c}$ without and with the shear-rotation-induced virialization term in the left and right panel, respectively. At high redshift, when matter perturbations are linear and dark energy perturbations negligible (they are in general about two orders-of-magnitude smaller than dark matter perturbations), the perturbed equation-of-state is very close to the background one, $w_\mathrm{c}\approx w_\mathrm{de}$. At later times, virialization alters its behaviour. 
When perturbations grow, the term $1/(1+\delta_\mathrm{de})$ becomes relevant; if virialization is not taken into account, all the models but the two phantom models DE2 and ODE have $w_\mathrm{c}=0$ at the collapse time. However, the time when the perturbed equation-of-state starts differing from the background equation-of-state depends on the model considered. Models DE1, DE2 and ODE deviate early as dark energy perturbations become more relevant at early times with respect to the other models. The CNR and 2EXP models, for example, start deviating at much later times. The models which deviate the earliest from the background value are the CPL and AS, since dark energy perturbations are more important. These results are expected in light of the previous discussion in section~\ref{sect:epsilon}.

The two phantom models have $w_{\rm c}\neq 0$ at the collapse time, and indeed $w_\mathrm{c}\to-\infty$. This divergence occurs because $w_\mathrm{c}=w_\mathrm{de}/(1+\delta_\mathrm{de})$ and for phantom models overdensities in matter correspond to underdensities in the dark energy component, yielding to a divergence in completely empty voids where $\delta_\mathrm{de}=-1$.

If we consider virialization induced by the presence of shear and rotation, neither $\delta_\mathrm{m}$ nor $\delta_\mathrm{de}$ become arbitrarily large, but reach a constant value, $\delta_\mathrm{m}\simeq\mathcal{O}(100)$. The perturbed equation-of-state $w_\mathrm{c}$ therefore does not vanish, as discussed above, but reaches a finite value. We show its time evolution as function of redshift in the right panel of figure~\ref{fig:wc} for the virialization model $S_1$, the difference with $S_2$ being negligible.

Recalling the discussion about the evolution of $\epsilon$, we easily understand that the two virialization recipes behave, at all effects, in the same way and differences in the final result are, in general, small (at most a few percent) and only relevant at late times when the density contrast is $\delta\gg 1$. As before, this general conclusion is true for quintessence models, but it does not apply to phantom models for the same reasons discussed before. In fact, as it is easy to see, the perturbed equation-of-state for the models DE2 and ODE becomes smaller than the background value. For DE2, the equation-of-state changes by a factor of 5 (7) for $S_1$ ($S_2$), while for ODE it changes by 70\% (80\%) for $S_1$ ($S_2$). Even if $\delta_\mathrm{de}\neq -1$, nevertheless $-1<\delta_\mathrm{de}<0$ and $|w_\mathrm{c}|>|w_\mathrm{de}|$.

Following the discussion before, we also understand that the models with a significant contribution from dark energy perturbations are, therefore, the CPL and the AS models. The other model with major deviations from the background value is the phantom model DE1, which goes from $-0.9$ to $\approx -0.4$.

\subsection{The virial overdensity \texorpdfstring{$\boldsymbol{\Delta_\mathrm{vir}}$}{DeltaV}}\label{sect:DeltaV}
In this section we present the main result of this work, the evolution of the virial overdensity $\Delta_\mathrm{vir}$. We compare the effects of dark energy perturbations on the models investigated with the $\Lambda$CDM model, which is our reference. Also for the $\Lambda$CDM model, we consider the two virialization recipes, solving Eqs.~(2.11) of \cite{Pace2019b}. For all models, both $\Lambda$CDM and dark energy, the virial overdensity represents the non-linear evolution of density perturbations $\Delta_\mathrm{vir} = 1+\delta_\mathrm{NL}$. We analyse two different cases: in the first we only consider matter perturbations, $\delta{_\mathrm{NL}}=\delta_\mathrm{m}$, while in the second we also explicitly take into account dark energy perturbations as defined in Eq.~(\ref{eqn:delta}) and discussed in \cite{Batista2017}.

\begin{figure}[!t]
 \centering
 \includegraphics[scale=0.4]{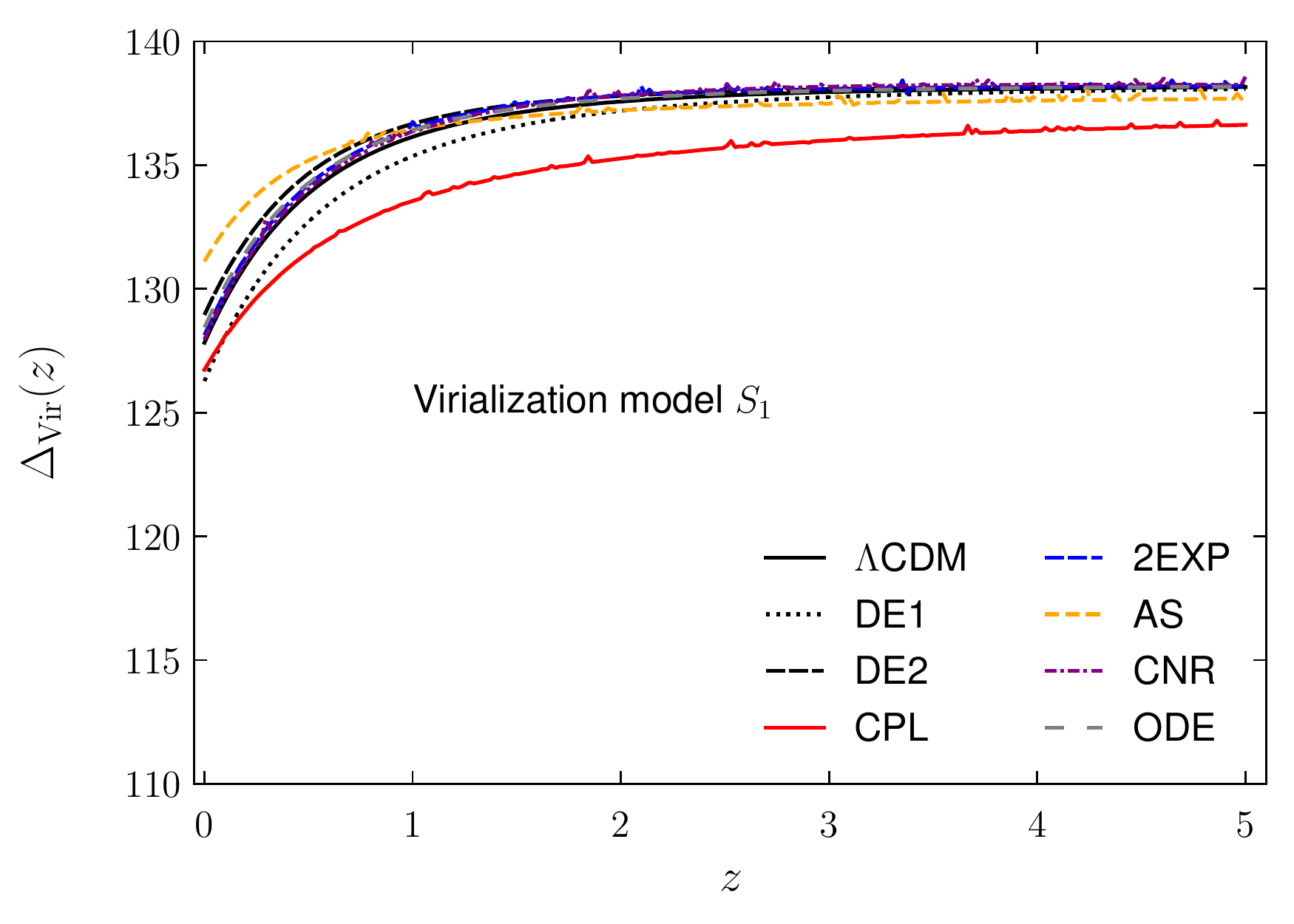}
 \includegraphics[scale=0.4]{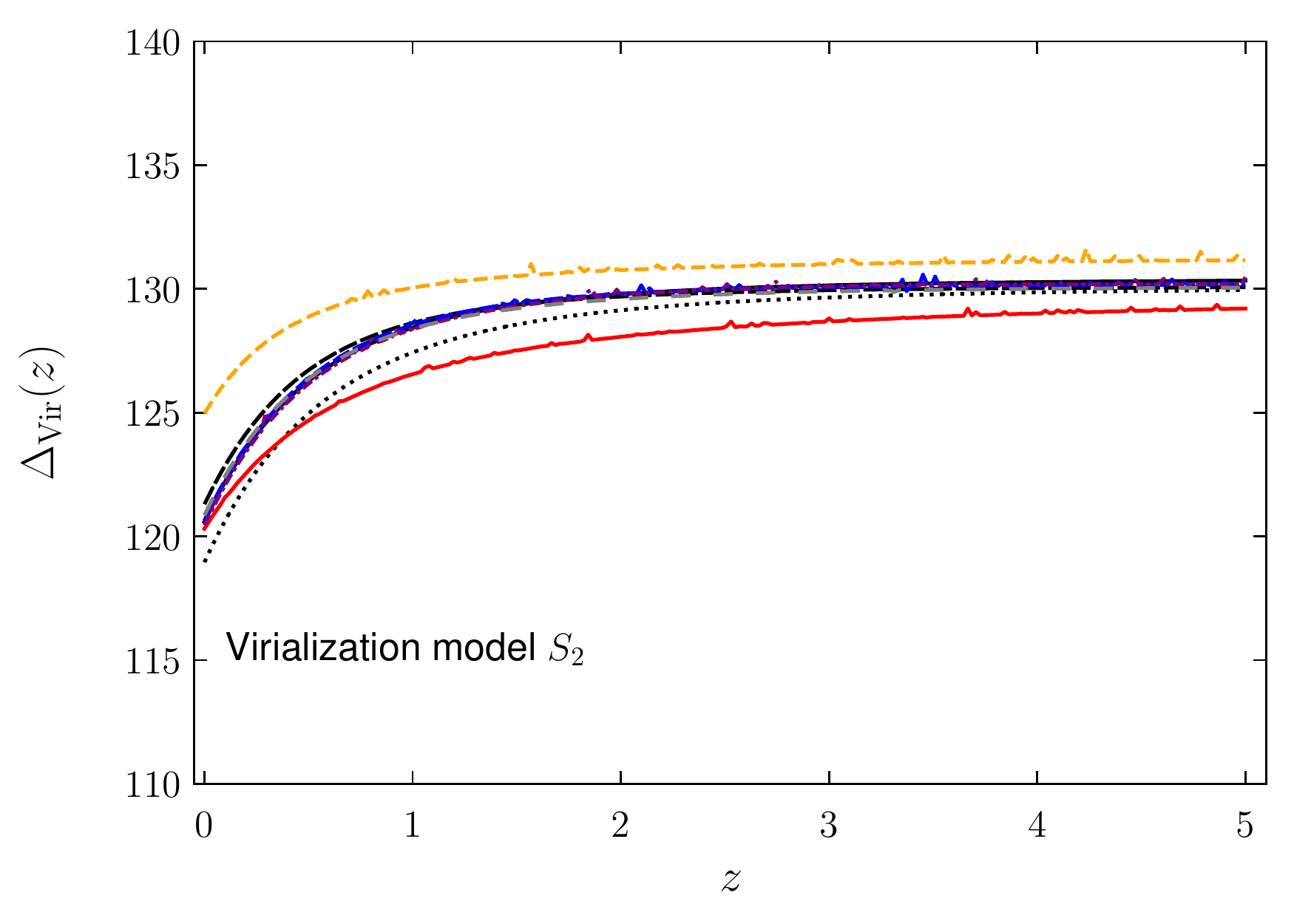}
 \includegraphics[scale=0.4]{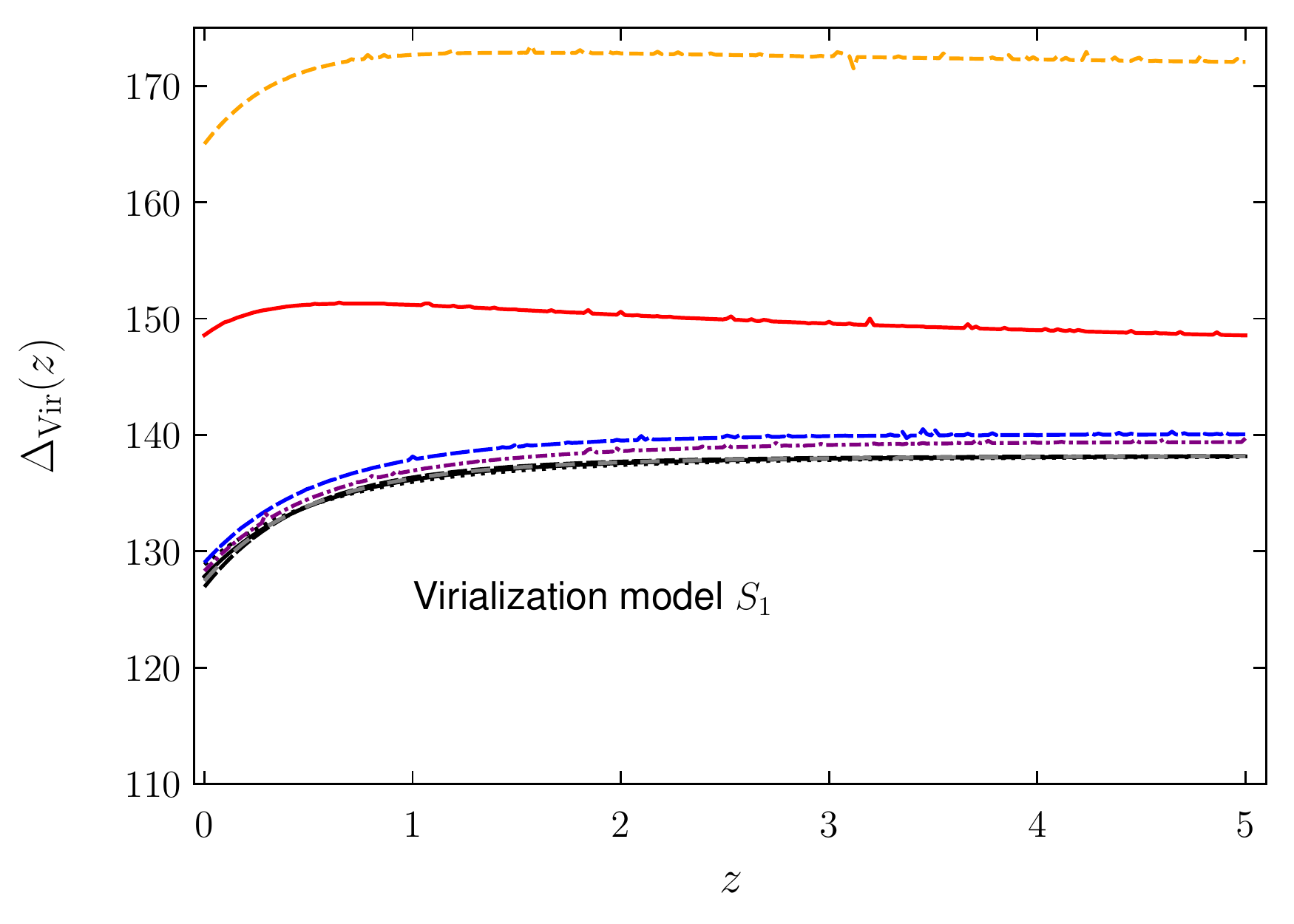}
 \includegraphics[scale=0.4]{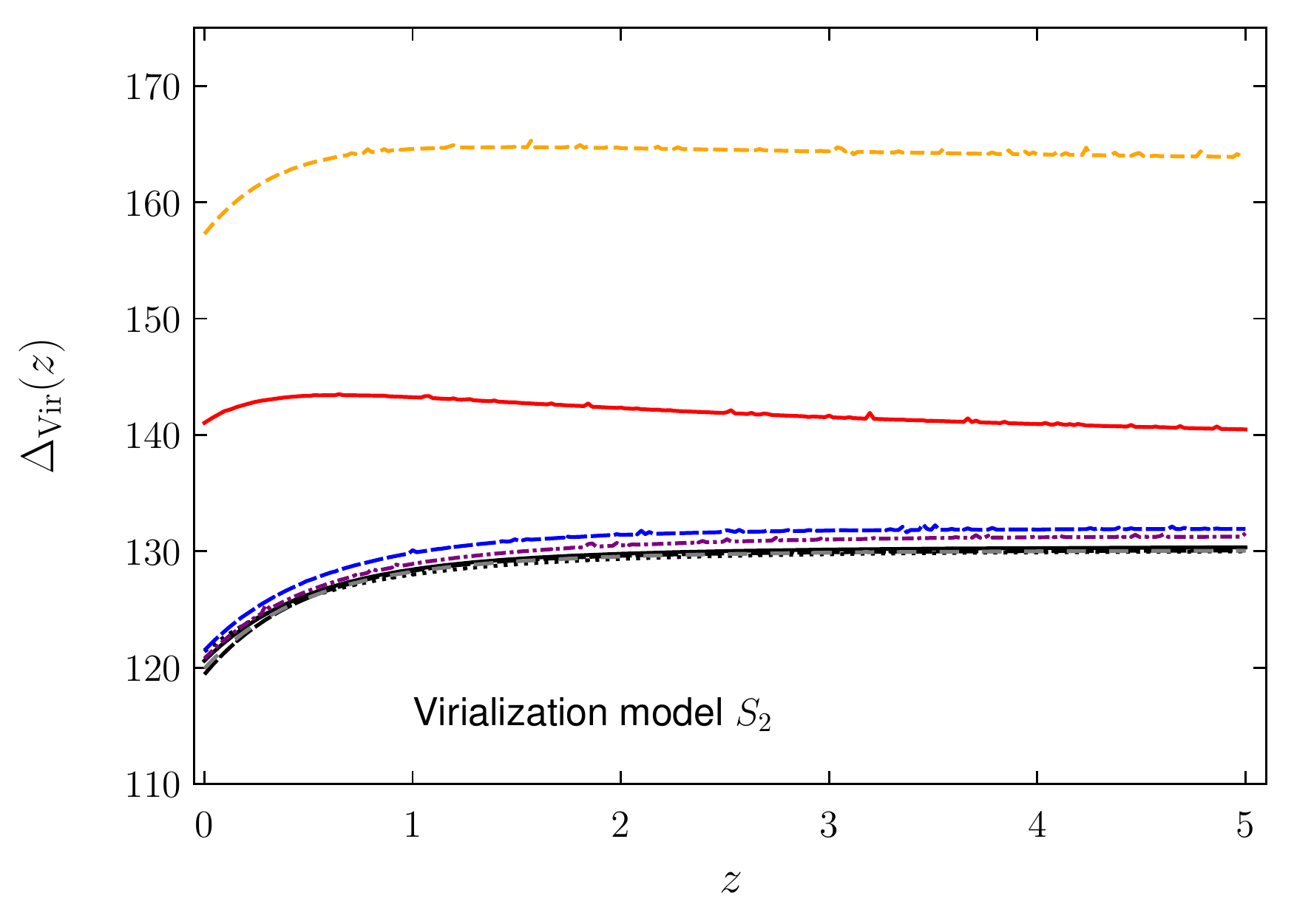}
 \caption{Time evolution of the virial overdensity $\Delta_\mathrm{vir}$ as a function of redshift $z$ for the $\Lambda$CDM model (black solid line) and the other dark energy models (see legend). Top (bottom) panels show the results ignoring (including) dark-energy fluctuations in the definition of $\delta$, namely $\delta=\delta_\mathrm{m}$ ($\delta=\delta_\mathrm{m}+\Omega_\mathrm{de}(a)\delta_\mathrm{de}/\Omega_\mathrm{m}(a)$). 
 Left (right) panels show the virialization model $S_1$ ($S_2$).}
 \label{fig:DeltaVir}
\end{figure}

In figure~\ref{fig:DeltaVir} we show the results of our analysis. In the upper panels we present the non-linear evolution of matter perturbations only (which are, though, affected by dark energy perturbations) while in the bottom panels we use a more general definition which makes explicit use of dark energy perturbations. For comparison, for the Einstein-DeSitter model the virial overdensity at virialization is time-independent and equal to $\Delta_{\rm vir}=18\pi^2(\tfrac{3}{4}+\tfrac{1}{2\pi})^2\simeq 147$.

The two recipes $S_1$ and $S_2$ give qualitatively the same behaviour either when only matter perturbations are considered or when dark energy perturbations are also explicitly taken into account. Focusing on the upper panels, we notice that all the models have very similar values of $\Delta_\mathrm{vir}$, except for CPL that shows a lower value. At high redshifts, the dark energy models but CPL converge to the same value, which is lower than for the $\Lambda$CDM of about one percent. The CPL instead, differs by about 2\% from the $\Lambda$CDM result. As it happens for the smooth dark energy case, prescription $S_2$ predicts lower values for $\Delta_\mathrm{vir}$, of about 5\%. This happens for the same reasons discussed for the smooth dark energy case, where $S_1(\delta)>S_2(\delta)$, as this condition still takes place for clustering dark-energy models on the regime of interest.

In the bottom panels, we explicitly take into account the contribution of dark energy. Again, we observe a very similar qualitative behaviour for the two prescriptions, with values for $S_2$ lower than for $S_1$. Two important differences occur with respect to the case where $\Delta_\mathrm{vir}=1+\delta_\mathrm{m}$. First, $\Delta_\mathrm{vir}$ for models DE2 and ODE is now $1-2$\% lower as $\delta_\mathrm{de}<0$. Second, there is a huge difference for both models CPL and AS with respect to top panels. This can be easily explained looking at figure~\ref{fig:epsilon}, which shows that for these models the dark energy contribution can be as large as 20\%. For example, at $z=0$ and for $S_1$ ($S_2$), $\Delta_\mathrm{vir}$ for the AS model rises from $\simeq 130$ (124) to $164$ (156). 
Note also that $\Delta_\mathrm{vir}$ for phantom (quintessence) models is smaller (larger) than for the $\Lambda$CDM model.

With respect to the case of smooth dark energy (see figure~6 of \cite{Pace2019b}), we limit ourselves to compare the upper panels only. Once again, qualitatively smooth and clustering dark-energy models have approximately the same behaviour and values do not differ much in the two cases. Major differences appear, as expected, for the CPL and AS models, being the contribution of dark energy more important than for the other models.

\section{Observables}\label{sect:observables}
We devote this section to study the impact of clustering dark energy on galaxy clusters counts, actually to counts of matter density peaks measured in gravitational lensing convergence maps, thermal Sunyaev-Zel'dovich (SZ) cosmic microwave background (CMB) maps, and in X-ray maps. Counts are the easiest cosmological observables controlled by the virialization mechanism able to probe the possible impact of dark energy fluctuations. We focus here on pure gravitational effects and ignore the baryonic physics apart from its role in calibration of mass-observable relations. Our target are forthcoming weak-lensing, CMB, and X-ray surveys, such as those operated by the Vera~Rubin~Observatory (VRO-LSST) \cite{Ivezic2019}, Euclid satellite \cite{Laureijs2009,Laureijs2011,Amendola2013}, Simons Observatory \cite{Hensley2021}, and extended Roentgen Survey with an Imaging Telescope Array (eROSITA) on the Spectrum-R\"ontgen-Gamma (SRG) satellite \cite{Predehl2010a,Predehl2010b,Merloni2012}.

\subsection{Virial mass and mass function}

\begin{figure}
 \centering
 \includegraphics[scale=0.4]{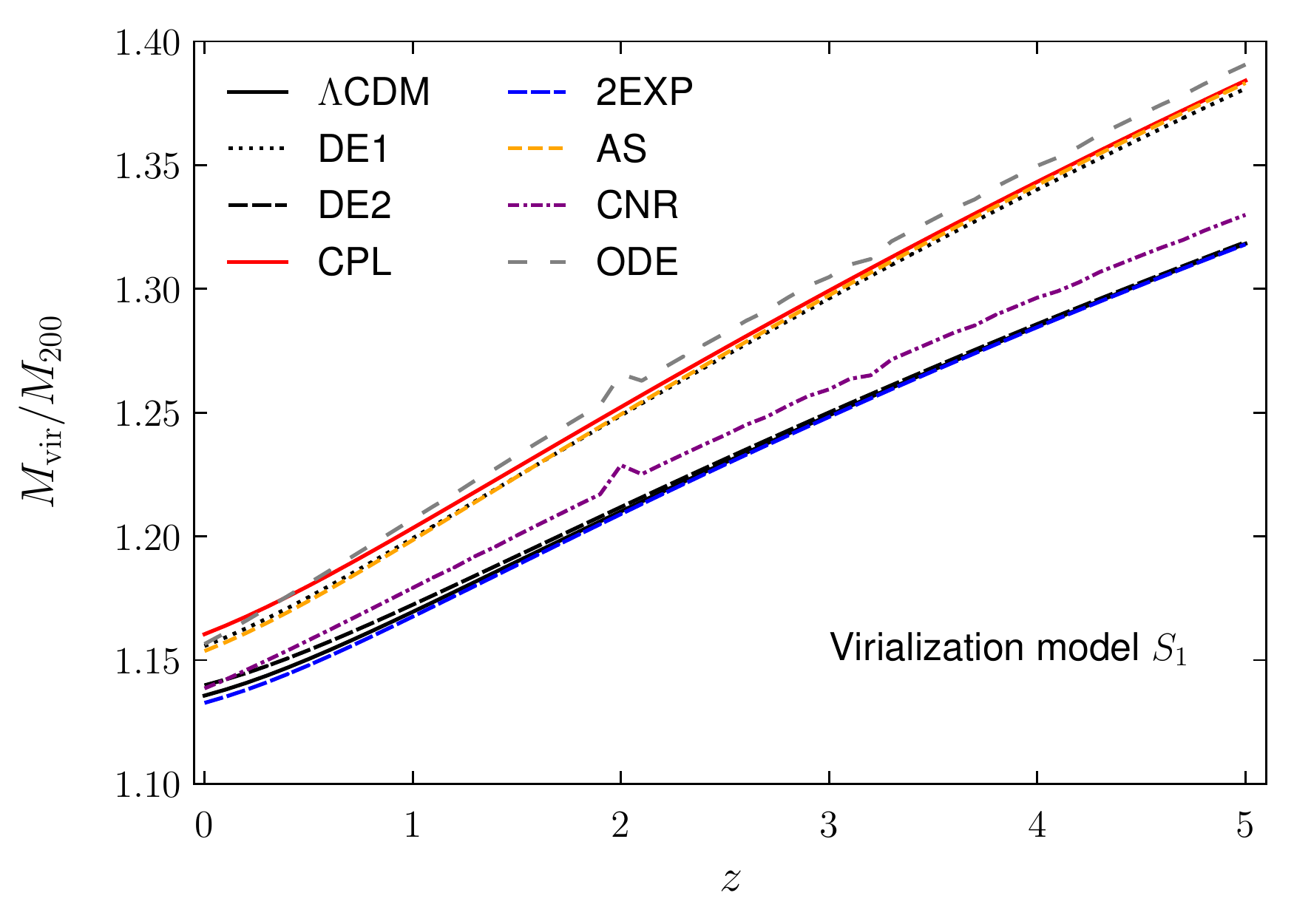}
 \includegraphics[scale=0.4]{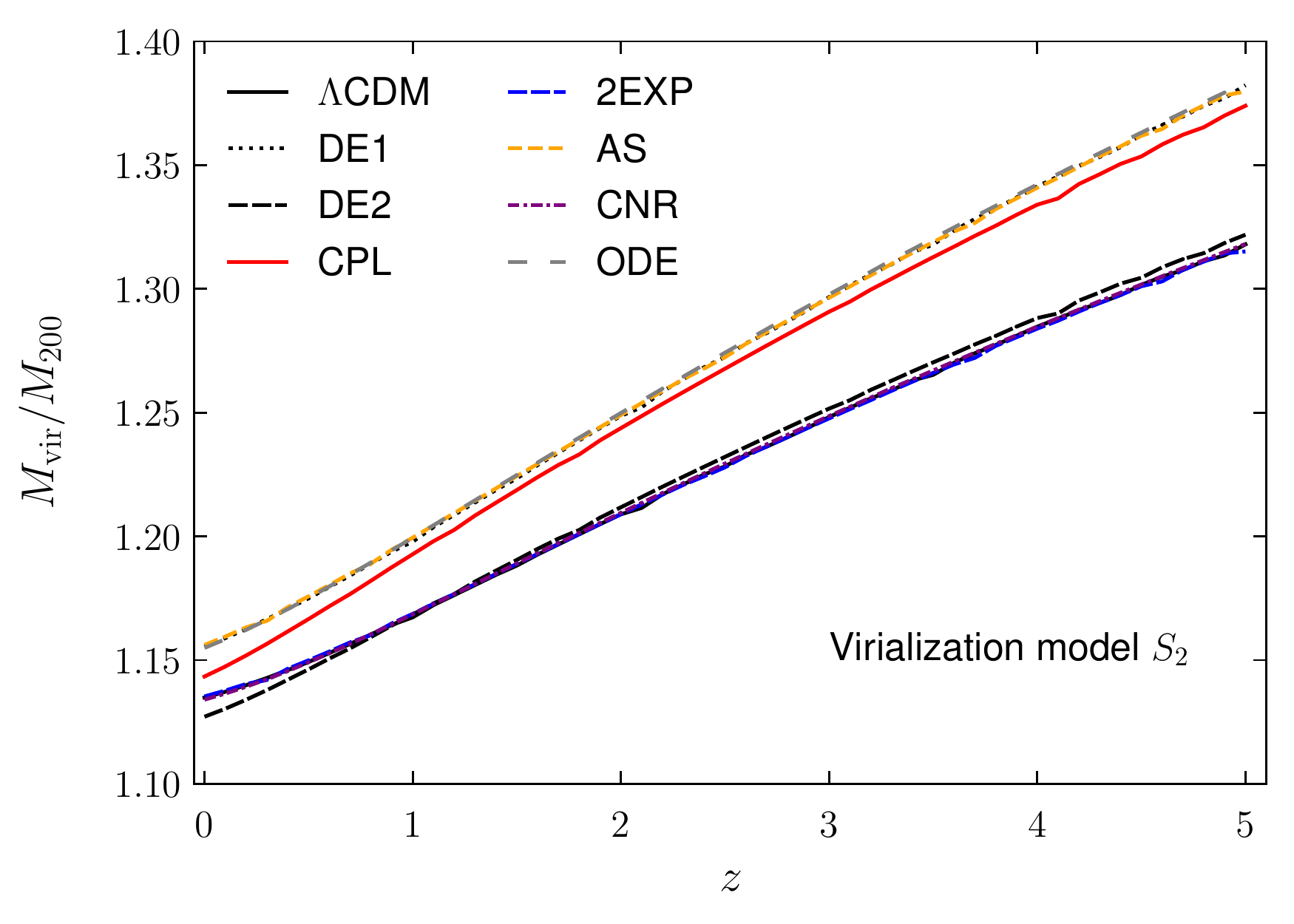}
 \includegraphics[scale=0.4]{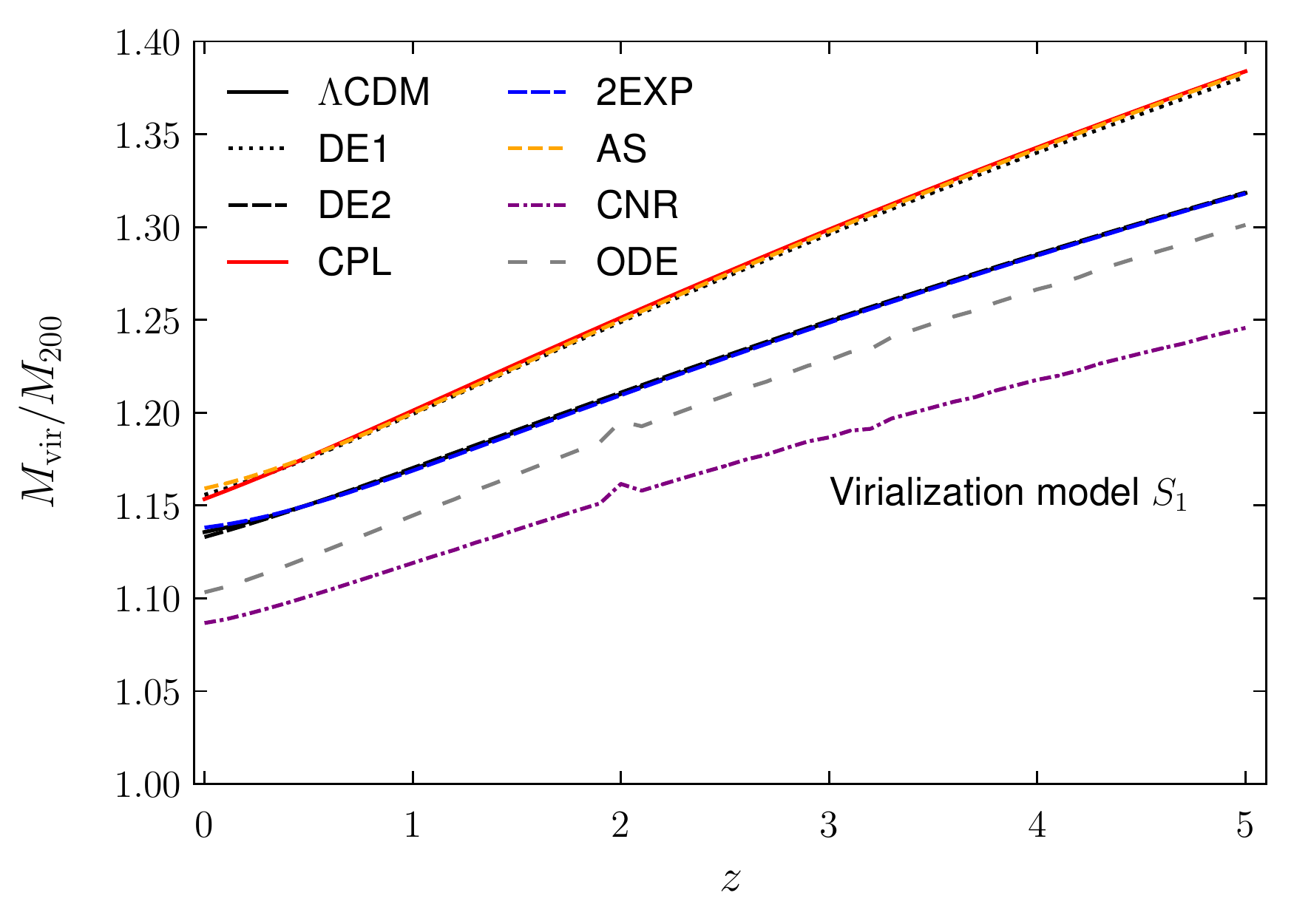}
 \includegraphics[scale=0.4]{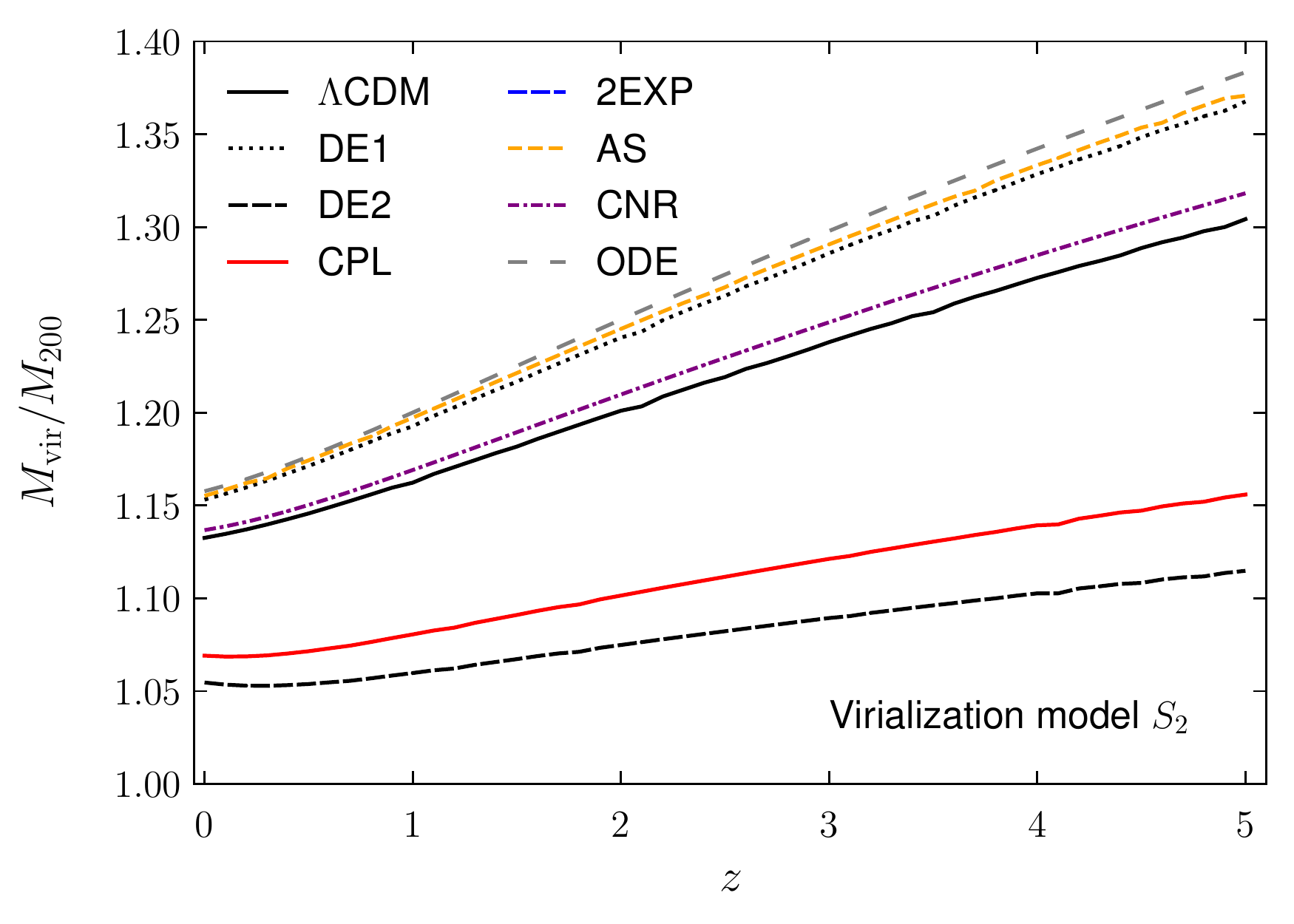}
 \caption{$M_\mathrm{vir}/M_{200}$ as a function of redshift for a halo with mass $M=10^{14.5}h^{-1}M_\odot$ and NFW profile, neglecting or including the dark-energy fluctuations (top panel and bottom panels, respectively) in the definition of the non-linear overdensity and for the $S_1$ and $S_2$ models (left and right panels, respectively).}
 \label{fig:Mvir}
\end{figure}

In the rest of the paper we will assume dark matter haloes with a truncated Navarro-Frank-White profile \cite{Takada2003b} and concentration parameter $c=c(M,z)$ modelled as in \cite{DeBoni2013}. For fixed dark-energy and virialization model, the relation between the virial mass $M_\mathrm{vir}$ and the total mass $M_{200}$ enclosed in a spherical halo with radius $R_{200}$ and density 200 times the critical density $\rho_\mathrm{cr}(z)=3H^2(z)/8\pi G$ is obtained solving the equation $f(u/c) u^3 = 200/\Delta_\mathrm{vir}(z) f(1/c)$ for $u=R_\mathrm{vir}/R_{200}$, with $f(x)=\log(1 + 1/x) - 1/(1 + x)$ for every value of mass and redshift \cite{Hu2003}, finally obtaining $M_{200}/M_\mathrm{vir}= 200/u^3\Delta_\mathrm{vir}(z)$. The same recipe is used for the $M_\mathrm{vir}-M_{500}$ relation. Figure~\ref{fig:Mvir} illustrates the results obtained for $M_\mathrm{vir}=10^{14.5}h^{-1}M_\odot$, which differs only in amplitude when considering other values of the virial mass. It is important to note that we obtained very similar results with an Einasto profile \cite{Gao2008,Navarro2010} (not shown); the overall amplitude of the WL and SZ signals and peak counts only slightly changes, and the relative differences between the clustering dark-energy models and the reference $\Lambda$CDM model is almost unchanged.

We employ the Watson mass function $n(M)$ \cite{Watson2013} with overdensity-dependent parametrization $\Delta$ (see their Equations~17-19), further multiplied by a factor $(1 + M_\mathrm{de}/M_\mathrm{m})$ to account for the dark-energy fluctuations. This correction factor modifies the amplitude of $n(M)$ by only few percents at redshift $z<1$ and does not have any effect at larger redshift for all but the CPL and AS models, whose fluctuations can instead imply variations of 5-10\% at redshift $z\simeq5$ and 15-20\% at $z=0$ depending on the value of their parameters. Apart from these extreme models, the mass function typically changes by less than one percent with respect to $\Lambda$CDM for all clustering dark-energy models and regardless of the virialization model.

Note that tidal virialization mainly affects the high-mass end of the halo mass function, reducing the number of the more massive haloes because of asphericity. This result was already present in the early study by \citep{Sheth2001} and can emerge also in anisotropic collapse models based, e.g., on a Bianchi metric of spacetime \citep{Giani2021}.

The complementary cumulative distribution function or tail distribution of lensing convergence peaks, SZ, and X-ray clusters observed with signal-to-noise ratio (SNR) or photon counts larger than $\rho$ reads
\begin{equation}\label{eq:Mcounts}
 N(>\rho) = \int \mathrm{d}\Omega \int \mathrm{d}z \int \mathrm{d}M\,\mathcal{S}(M,z,\rho,l,b)\frac{\mathrm{d}V}{\mathrm{d}z\mathrm{d}\Omega}n(M)\,.
\end{equation}
Here $\mathrm{d}\Omega$ is the solid angle element, $\mathrm{d}V/\mathrm{d}z\mathrm{d}\Omega$ is the differential comoving volume, and $\mathcal{S}$ is the cluster selection function for the survey, which depends on the mass $M$, redshift $z$ and (galactic) angular coordinates $(l,b)$ of clusters, and on $\rho$. The mass-observable relation, the intrinsic noise of the observable, the projected morphology of clusters, and the angular resolution of the instrument define the functional form of $\rho$.

\subsection{Convergence peaks: forecasts for VRO-LSST and Euclid-like surveys}
 
The estimation follows the standard operations described in \cite{Pace2019b}. For fixed dark-energy and virialization models, each spherical dark-matter halo at redshift $z$ with virial overdensity $\Delta_\mathrm{vir}(z)$, truncated Navarro-Frank-White profile \cite{Takada2003b}, and corresponding virial mass $M$ yields a radial surface-mass-density $\Sigma(R)$ as function of the coming halo radius $R$.

Such a halo located at angular distance $d_{\rm A}(z)$ from the observer, operating as lens of background sources distributed according to a redshift distribution parametrised by $p(z_s)=(z_s/z_0)^\alpha\exp{[-(z_s/z_0)^\beta]}$ with $\alpha$ and $\beta$ specific to the survey, produces a convergence $\kappa=\Sigma(R)/\Sigma_\mathrm{cr}$. The mean critical surface-mass-density $\Sigma_\mathrm{cr}$ depends on the background cosmology and on $p(z_s)$ \cite{Narayan1996,Bartelmann2001}.

\begin{figure}
 \centering
 \includegraphics[scale=0.4]{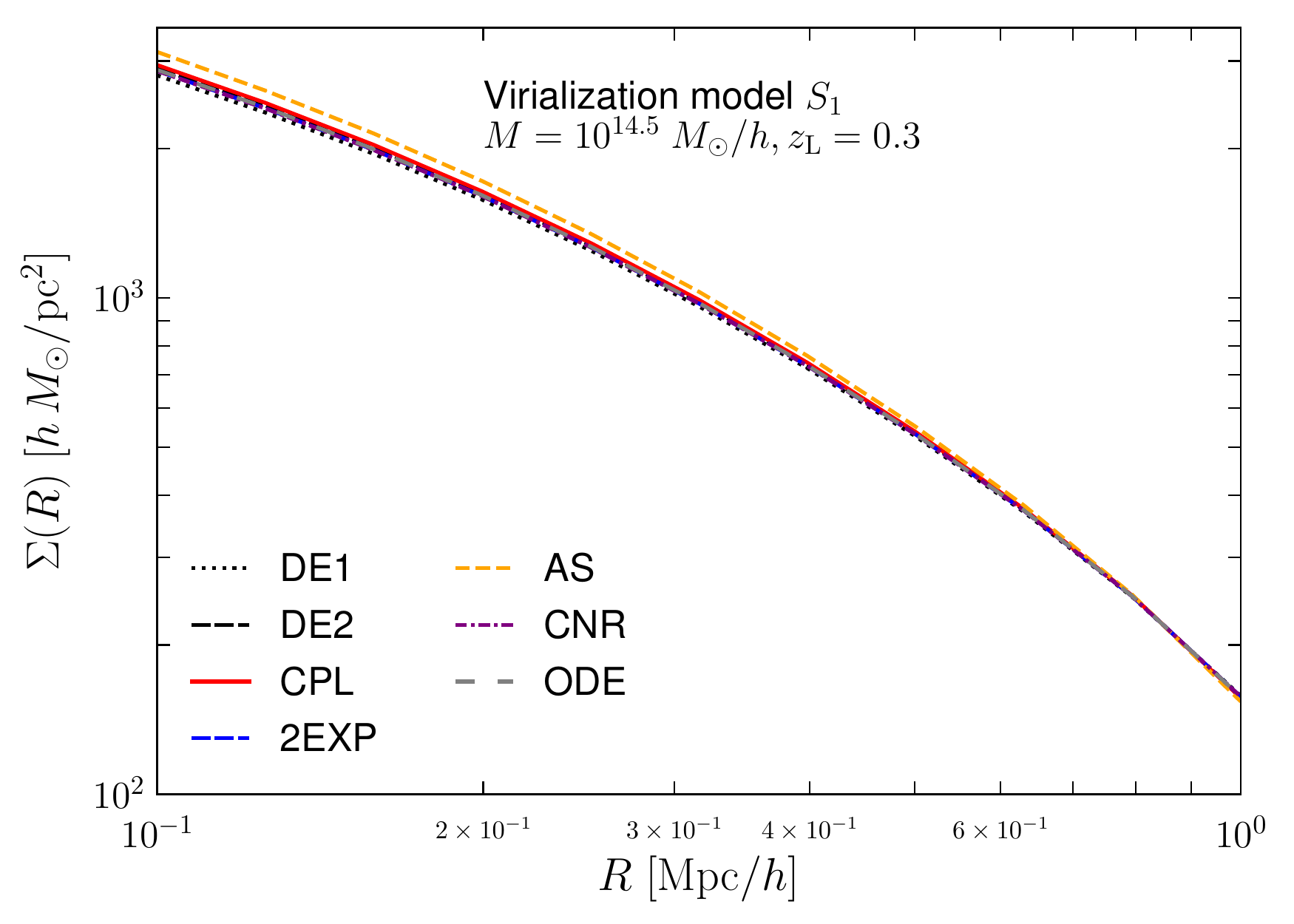}
 \includegraphics[scale=0.4]{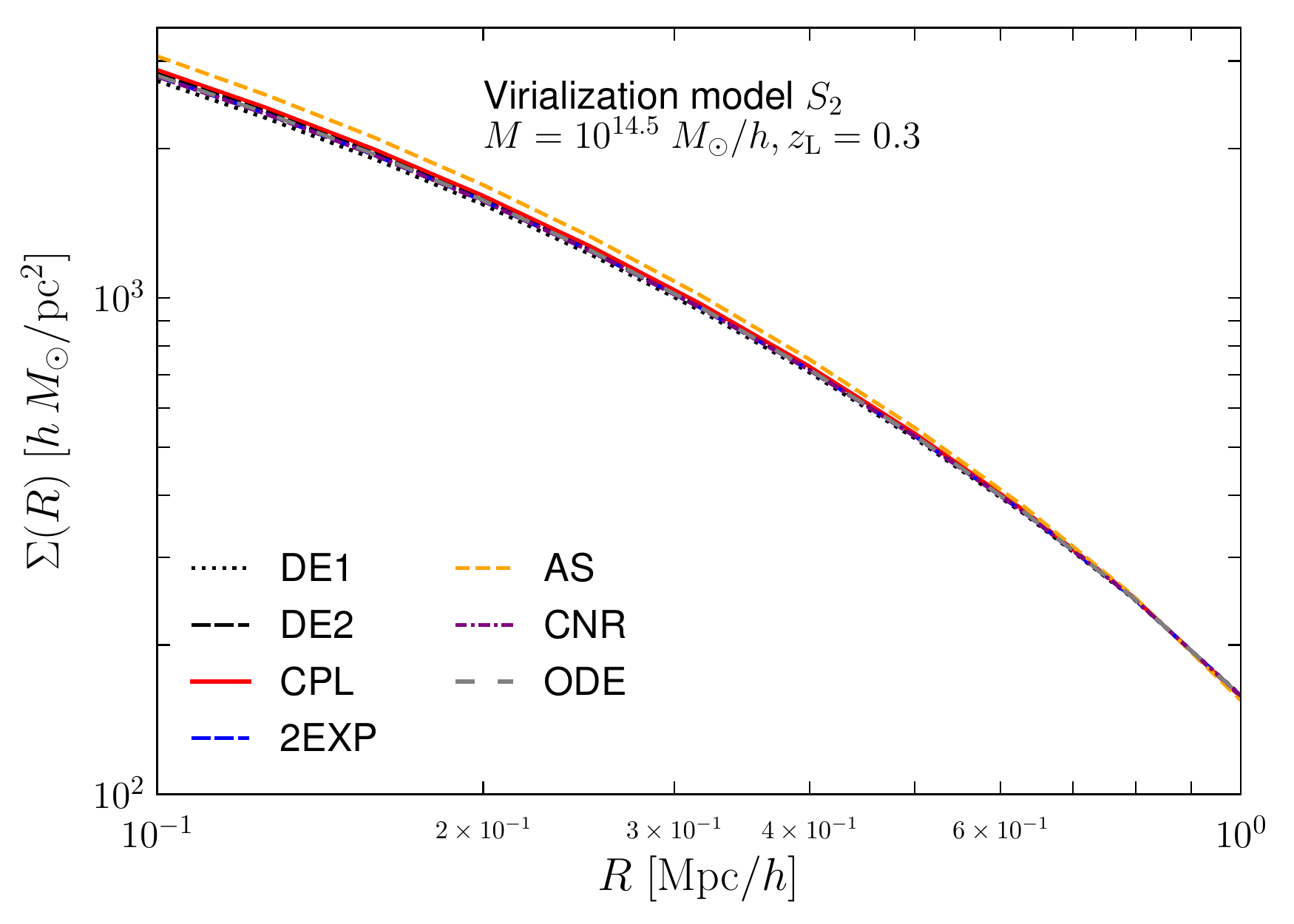}
 \includegraphics[scale=0.4]{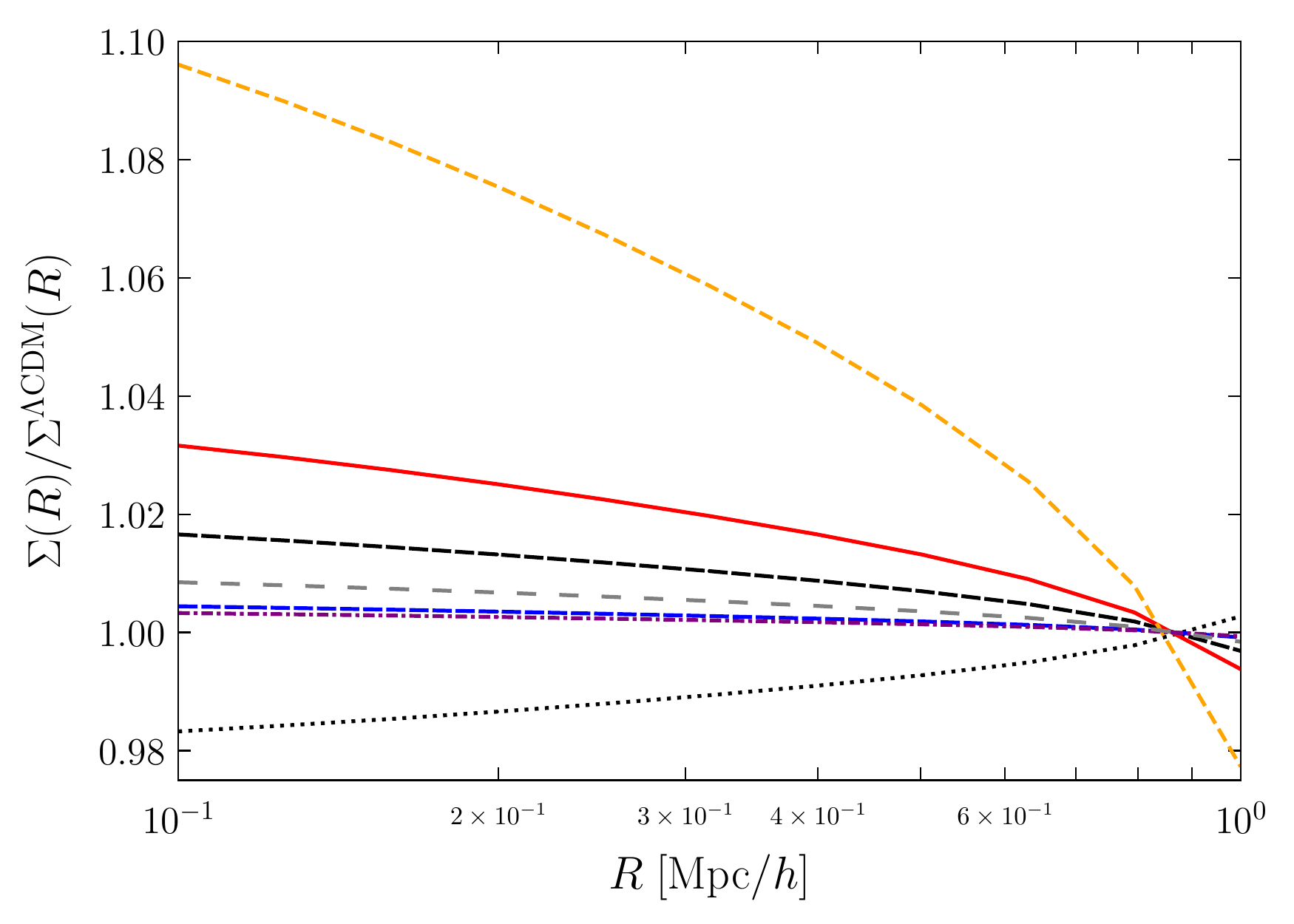}
 \includegraphics[scale=0.4]{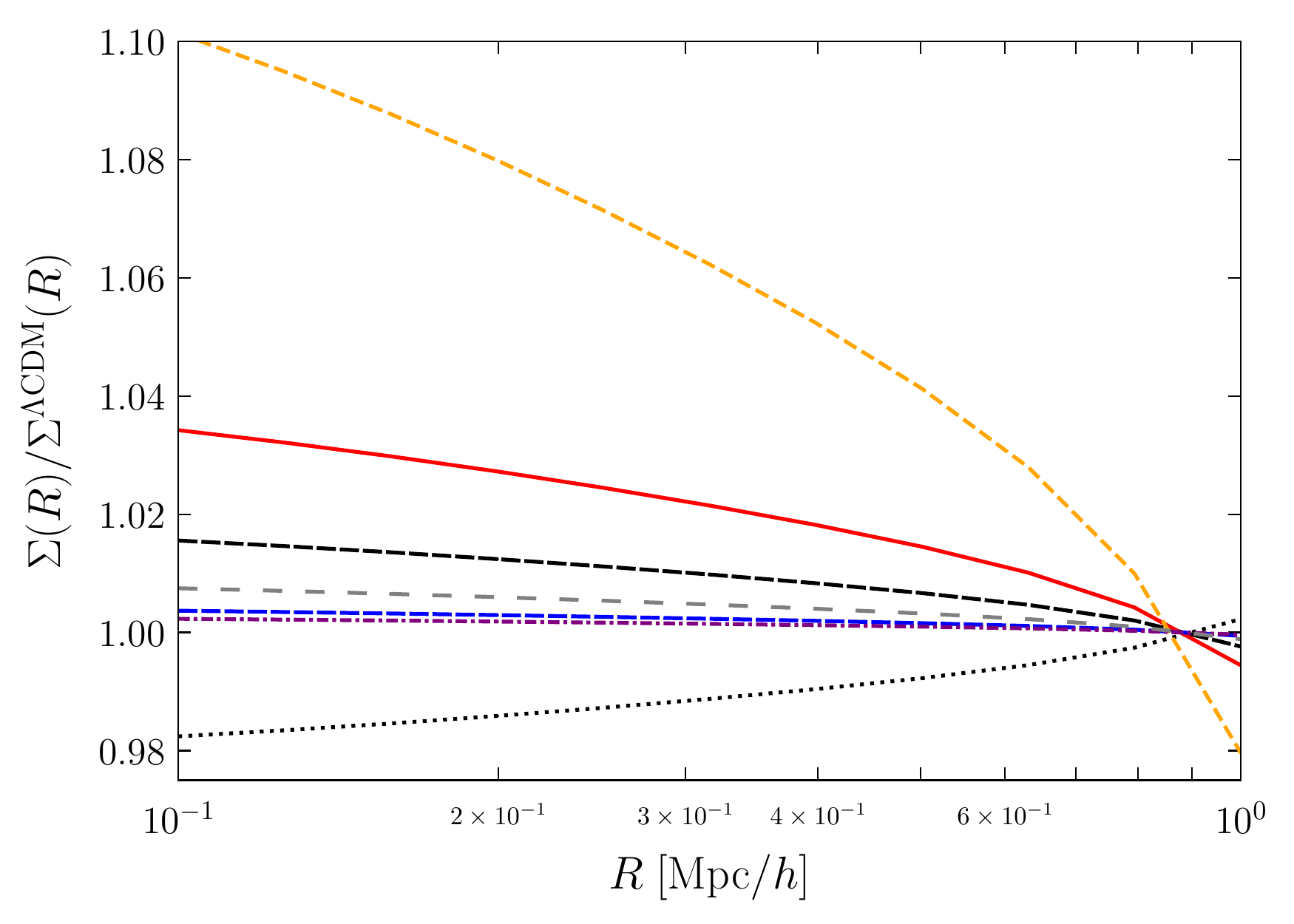}
 \includegraphics[scale=0.4]{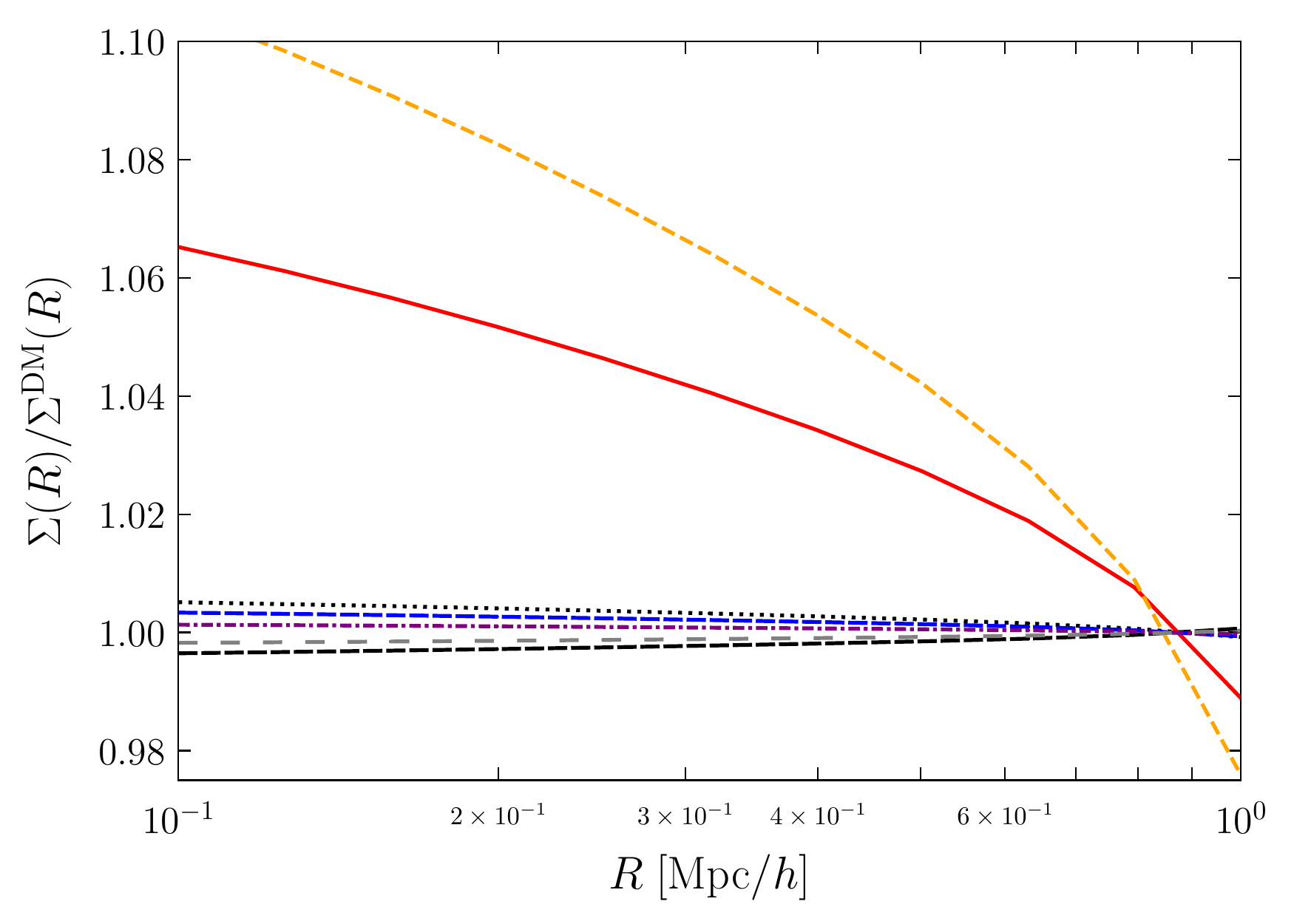}
 \includegraphics[scale=0.4]{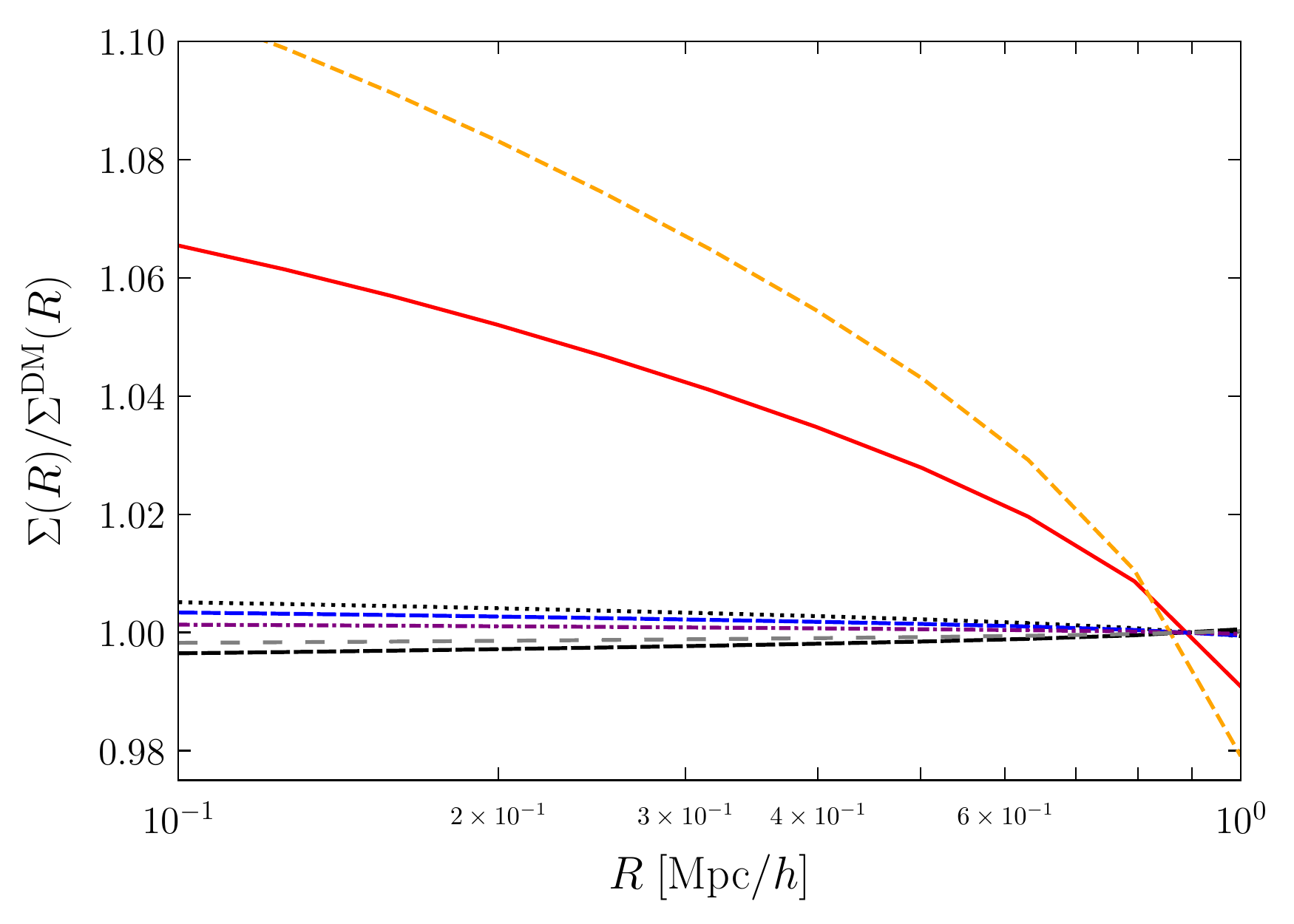}
 \caption{
 Radial surface-mass-density $\Sigma(R)$ for a halo of mass $M=10^{14.5}\,h^{-1}M_\odot$ at redshift $z_\mathrm{L}=0.3$ with truncated NFW profile, for the two virialization recipes $S_1$ (left panels) and $S_2$ (right panels) and the seven dark energy models (line/colour styles as in figure~\ref{fig:DeltaVir}).
 Top panels: absolute values. Middle and bottom panels: absolute variations w.r.t. $\Lambda$CDM model and smooth dark-energy models.}
 \label{fig:Sigma1}
\end{figure}

Figure~\ref{fig:Sigma1} illustrates the result for a halo with mass $M=10^{14.5}h^{-1}M_\odot$ at redshift $z=0.3$ for the reference $\Lambda$CDM cosmology and dark energy models (top panels), the relative deviations for the dynamical dark-energy models with fluctuations with respect to the $\Lambda$CDM (central panels), and relative deviations for the dynamical dark-energy models with fluctuations with respect to the same models without fluctuations, to illustrate the effect of fluctuations (bottom panels). All the results are shown for the two virialization models, which differ for a very tiny amount unlike the intrinsic difference due to the dark-energy model. Qualitative very similar results for deviations are obtained looking at the convergence field or considering haloes at different redshift.

\begin{table}
 \caption{Relative deviation of number counts of convergence peaks with signal-to-noise $\nu > 3$ (upper table) and $\nu > 5$ (lower table) w.r.t. $\Lambda$CDM cosmology for VRO-LSST and Euclid-like weak lensing surveys. Quoted in parenthesis, relative deviation w.r.t. corresponding smooth dark-energy model, i.e., without perturbations ($\delta_\mathrm{de}=0$). As reference, absolute number counts with Poisson error are reported for the $\Lambda$CDM model. Values are stable against reasonable variations of cosmological parameters defining the reference model.
 }
 \begin{center}
 \resizebox{0.98\hsize}{!}{
  \begin{tabular}{|c|cc||cc|}
   \hline
   & \multicolumn{2}{c||}{VRO-LSST-like} & \multicolumn{2}{c|}{Euclid-like}\\
   \cline{2-5}
    & $S_1$ & $S_2$ & $S_1$ & $S_2$ \\
   \hline\hline
   & \multicolumn{4}{c|}{signal-to-noise $\nu > 3$} \\
   \hline
   $\Lambda$CDM & $32843\pm181$ ($0.5\%$) & $31238\pm177$ ($0.6\%$) & $36078\pm190$ ($0.6\%$) & $34700\pm186$ ($0.6\%$) \\
   \hline
   DE1 & $-10.3\%$ ($+1.3\%$) & $-10.8\%$ ($+1.4\%$) & $-9.1\%$ ($+1.6\%$) & $-9.5\%$ ($+1.6\%$) \\
   DE2 &  $+10.1\%$ ($-0.7\%$) & $+10.3\%$ ($-0.7\%$) &  $+9.2\%$ ($-0.8\%$) & $+9.3\%$ ($-0.9\%$) \\
   CPL & $-13.3\%$ ($+30.8\%$) & $-13.1\%$ ($+31.7\%$) &  $-5.8\%$  ($+28.7\%$) & $-5.5\%$ ($+29.5\%$) \\
   2EXP & $+1.2\%$ ($+1.6\%$) & $+1.0\%$ ($+1.7\%$) &  $+1.1\%$ ($+1.4\%$) & $+1.0\%$ ($+1.5\%$) \\
   AS & $+5.1\%$ ($+55.8\%$) & $+6.4\%$ ($+57.2\%$) & $+13.1\%$ ($+49.8\%$) & $+14.5\%$ ($+51.0\%$) \\
   CNR & $+0.5\%$ ($+0.6\%$) & $+0.3\%$ ($+0.7\%$) & $+0.5\%$ ($+0.6\%$) & $+0.4\%$ ($+0.6\%$) \\
   ODE & $+5.0\%$ ($-0.3\%$) & $+5.0\%$ ($-0.3\%$) & $+4.6\%$ ($-0.4\%$) & $+4.5\%$ ($-0.4\%$) \\
   \hline\hline
   & \multicolumn{4}{c|}{signal-to-noise $\nu > 5$} \\
   \hline
   $\Lambda$CDM & $1482\pm39$ ($2.6\%$) & $1316\pm36$ ($13.3\%$) & $2652\pm52$ ($2.2\%$) & $2532\pm50$ ($2.2\%$) \\
   \hline
   DE1 & $-16.4\%$ ($+2.7\%$) & $-17.3\%$ ($+2.8\%$) & $-8.7\%$ ($+1.8\%$) & $-7.7\%$ ($+2.3\%$) \\
   DE2 &  $+17.1\%$ ($-1.4\%$) & $+1.2\%$ ($-1.5\%$) & $+9.8\%$ ($-1.3\%$) & $+8.1\%$ ($-1.5\%$) \\
   CPL & $-17.4\%$ ($+56.5\%$) & $-16.7\%$ ($+58.7\%$) & $-17.4\%$  ($-36.8\%$) & $-21.0\%$ ($-41.4\%$) \\
   2EXP & $+2.1\%$ ($+2.6\%$) & $+1.8\%$ ($+2.7\%$) & $-0.5\%$ ($+1.5\%$) & $+0.9\%$ ($+1.1\%$) \\
   AS & $+14.8\%$ ($+104.8\%$) & $+18.3\%$ ($+108.1\%$) & $+24.1\%$ ($+55.7\%$) & $+24.7\%$ ($+52.2\%$) \\
   CNR & $+0.9\%$ ($+1.0\%$) & $+0.5\%$ ($+1.1\%$) & $+0.3\%$ ($+0.8\%$) & $+0.7\%$ ($+0.4\%$) \\
   ODE & $+8.4\%$ ($-0.7\%$) & $+8.2\%$ ($-0.8\%$) & $+4.5\%$ ($-0.4\%$) & $+6.3\%$ ($-0.1\%$) \\
   \hline
  \end{tabular}
  }
  \label{tab:kappapeakcounts}
 \end{center}
\end{table}

Convergence peaks counts are deduced from the azimuthally averaged convergence $\kappa_G(\theta_G) = \int \mathrm{d}^2\boldsymbol{\theta}W(\boldsymbol{\theta};\theta_G)\kappa(\boldsymbol{\theta})$, which measures the average isotropic deformation of background images seen at angle $\theta_{\rm G}=R/d_{\rm A}(z)$ with respect to the centre of the halo, customarily coincident with the centre of the Gaussian smoothing filter $W(\boldsymbol{\theta};\theta_{\rm G})$. The number of averaged convergence peaks with signal-to-noise ratio $\rho\equiv\nu=\kappa_{\rm G}(M)/\sigma_\mathrm{noise}$ exceeding a fixed threshold $\nu_\mathrm{th}$ is finally computed integrating over the (comoving) survey volume $V$ and mass function $n(M)$, which yields
\begin{equation}\label{eq:WLcounts}
 N(\nu > \nu_\mathrm{th}) = \int \mathrm{d}z \int \mathrm{d}M \frac{\mathrm{d}V}{\mathrm{d}z}n(M)\mathcal{H}(\nu>\nu_\mathrm{th})\,.
\end{equation}
Here $\mathcal{H}(x)$ is the unit step function and the noise variance $\sigma_\mathrm{noise}^2=(\sigma_\epsilon^2/2)/2\pi\theta_{\rm G}^2n_\mathrm{g}$ is specific to the survey via the standard deviation of the intrinsic ellipticity distribution, $\sigma_\epsilon$, and the number of lensed galaxies per unit angle, $n_\mathrm{g}$. The parameters of the VRO-LSST and Euclid-like surveys are $(n_\mathrm{g}/\mathrm{arcmin}^2,\alpha,\beta,z_0,\sigma_\epsilon,\theta_\mathrm{G}) = (26,1.28,0.97,0.41,0.3,1.875)$ \cite{Chang2013} and $(35,2,1.8,0.7,0.3,2.35)$ \cite{Laureijs2009,Laureijs2011}, respectively. Both are supposed to cover an effective area of 15000~deg$^2$ resulting after masking and detect haloes in the redshift range $0.3<z<0.7$ (VRO-LSST-like) and $0.3<z<1.5$ (Euclid-like).

The results of counts for signal-to-noise ratio larger than $\nu_\mathrm{th}=3$ and 5 are summarised in table~\ref{tab:kappapeakcounts}, which shows the relative variation (ratio) w.r.t. the counts expected for the $\Lambda$CDM model and that for the corresponding smooth dark-energy model (values quoted in parenthesis). As absolute reference, we report the total number counts for the two surveys and an estimation of the Poissonian error and its relative value (in parenthesis). Despite the minor difference between the surface mass density (or average converge) profiles of single haloes, the depth and the large sky coverage of these weak-lensing surveys allow for an appreciable difference in counts between dark-energy and virialization models. Interestingly enough, at this level of approximation the impact of dark-energy fluctuations is not negligible, especially for CPL and AS models, opening the possibility to investigate non-trivial dynamical dark-energy models by future surveys. All but the 2EXP and CNR models might be potentially distinguished from $\Lambda$CDM already with signal-to-noise ratio as small as 3. Taking into account the absolute value of counts, the difference between the two virialization models $S_1$ and $S_2$ could also be appreciated, the difference between the number counts for the $\Lambda$CDM model being about $4.5\sigma$ to $6.5\sigma$ for signal-to-noise ratio $\nu>3$ and $1.5\sigma$ to $3\sigma$ for $\nu>5$; similar values are obtained for the other dark-energy models.

\subsection{SZ peaks: forecasts for Simons Observatory-like survey}

We model the thermal Sunyaev-Zeldovich (SZ) signal emitted by a halo following \cite{Planck2013_XX,Planck2015_XXIV}. The mass-observable relation between the angular-averaged Comptonization parameter $\bar{Y}_{500}$ and the total mass $M_{500}$ of a halo located at redshift $z$ is given by
\begin{equation}\label{eq:YMscaling}
 E^\beta(z)\left[\frac{d_A^2(z)\bar{Y}_{500}}{10^{-4}\mathrm{Mpc}}\right] = Y_{\ast} \left[\frac{h}{0.7}\right]^{-2+\alpha} \left[\frac{(1-b)M_{500}}{6\times10^{14}M_\odot}\right]\,,
\end{equation}
with $E(z)=H(z)/H_0$ and fiducial parameters $(\log Y_{\ast},\alpha,\beta)=(-0.19,1.79,0.66)$. This scaling is established by X-ray observations, which found a pretty linear relation between the hydrostatic equilibrium mass of the intergalactic medium and $M_{500}$ with slope $(1-b)=0.8$, customarily assumed constant both in mass and redshift. The calibration of the $Y-M$ scaling relation, which can be further tightened by gravitational lensing measurement based on CMB-Stage 4 experiments \cite{Madhavacheril2017,Louis2017}, is not discussed here and goes beyond the scope of this study.

The number of SZ haloes in the surveyed area and redshift range with signal-to-noise ratio\footnote{We adopt the same notation as Planck Collaboration papers, this quantity should, therefore, not be confused with the free parameter $q$ introduced for the $S_1$ and $S_2$ recipes.} $\rho\equiv q=Y_{500}/\sigma_\mathrm{noise}$ larger than $q_\mathrm{th}$ is given by
\begin{equation}\label{eq:SZcounts}
 N(q > q_\mathrm{th})=\int \mathrm{d}\Omega\int\mathrm{d}z\int \mathrm{d}M_{500}~\hat{\chi}(z,M_{500},l,b)\frac{\mathrm{d}V}{\mathrm{d}z}n(M_{500})\,,
\end{equation}
where
\begin{equation}\label{eq:SZmeancompleteness}
 \hat{\chi}(z,M_{500},l,b) = \int_{q_\mathrm{th}}^\infty \mathrm{d}q~P[q|\bar{q}(M_{500},z,l,b)] = \int\mathrm{d}\ln Y_{500}~P[\ln Y_{500}|z,M_{500}]\,\chi(Y_{500},l,b)\,,
\end{equation}
is the tail distribution of $q$ with mean value $\bar{q}$ given by Eq.~\eqref{eq:YMscaling}. The latter can also be interpreted as an averaged completeness integral of the survey selection function $\chi(Y_{500},l,b)$ at $q>q_\mathrm{th}$ over the probability of measuring $Y_{500}$ for a halo of mass $M_{500}$ located at galactic coordinates $(l,b)$. As usual, we assume the latter as a log-normal law with mean $\bar{Y}_{500}(M_{500},z)$ given by Eq.~(\ref{eq:YMscaling}) and $\sigma_{\ln Y}=0.173$. For the purpose of this study, it is sufficient to adopt the ERF model for the survey selection function \citep[see Equation~(14) in][]{Planck2015_XXIV} and a constant filter noise independent of the halo location and angular size $\theta_{500}=R_{500}/d_{\rm A}(z)$ matching the average Planck noise, $\sigma_\mathrm{noise}=\langle\sigma_Y\rangle = 2.2\times10^{-4}\mathrm{arcmin}^2$, or ten times smaller as might be expected for future surveys. We shall consider a SO-like survey with effective area covering 16500~deg$^2$, i.e., $0.4$ times the full sky. More accurate estimation goes beyond the scope of this study.

\begin{figure}[!t]
 \centering
 \includegraphics[scale=0.4]{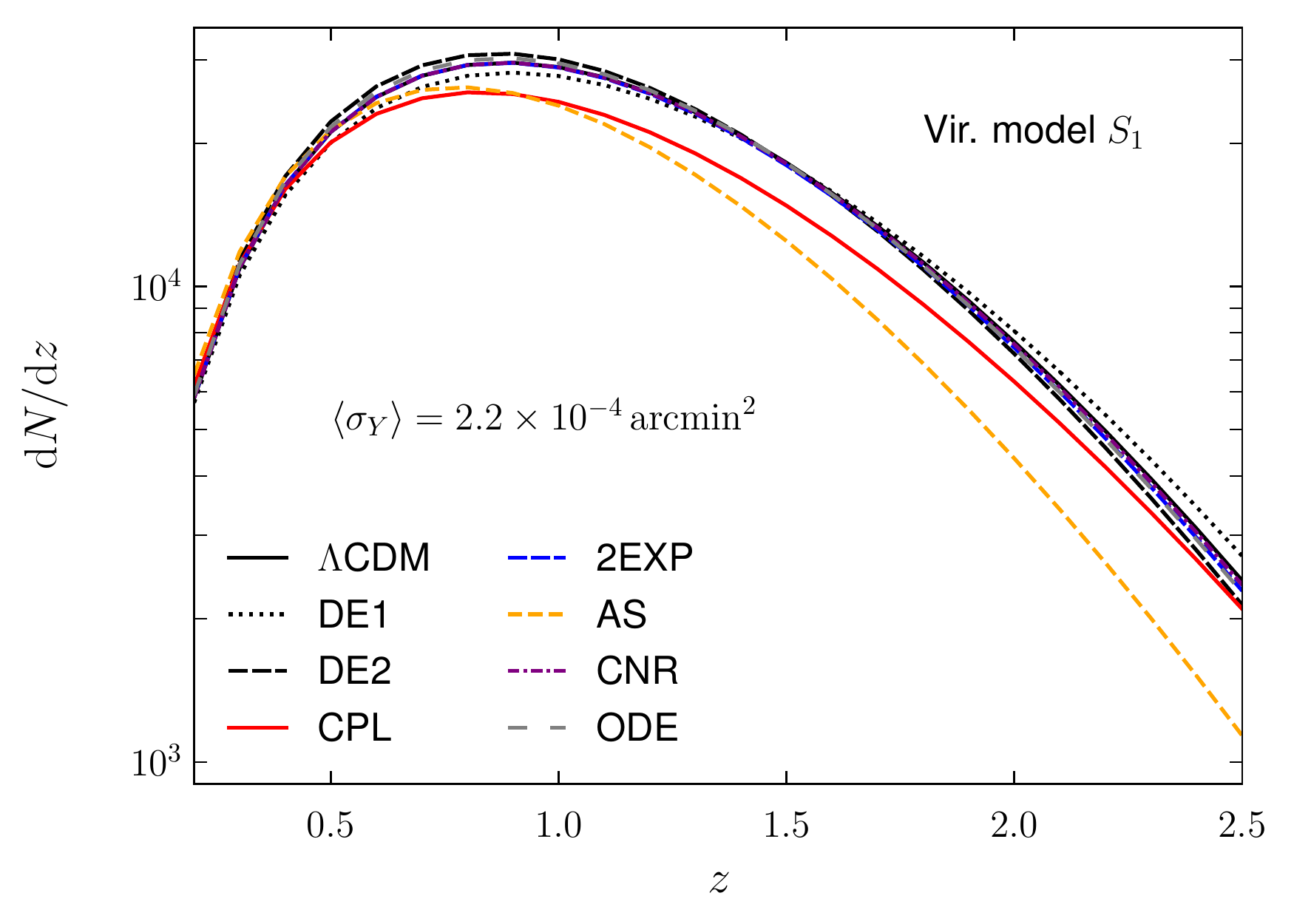}
 \includegraphics[scale=0.4]{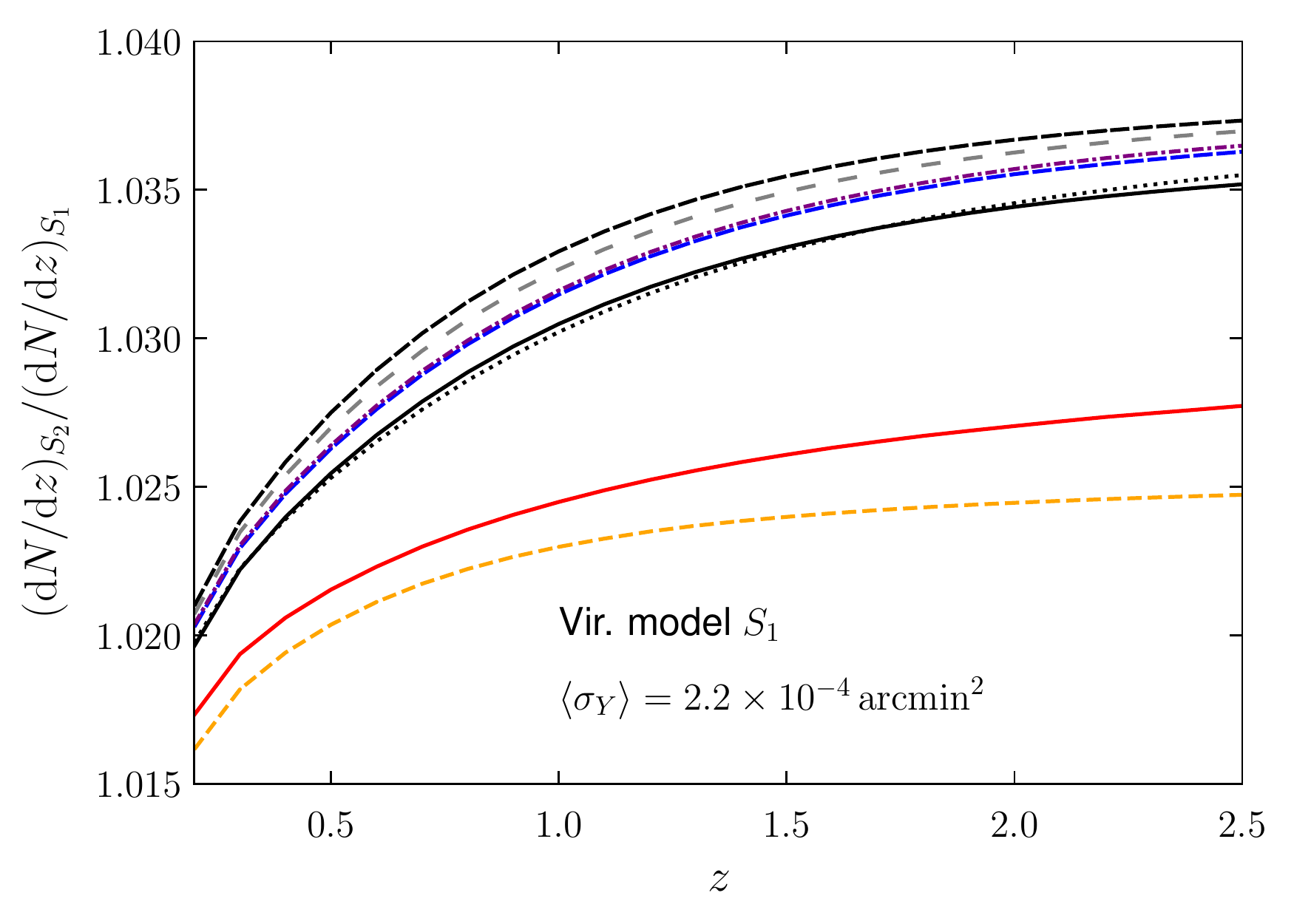}
 \includegraphics[scale=0.4]{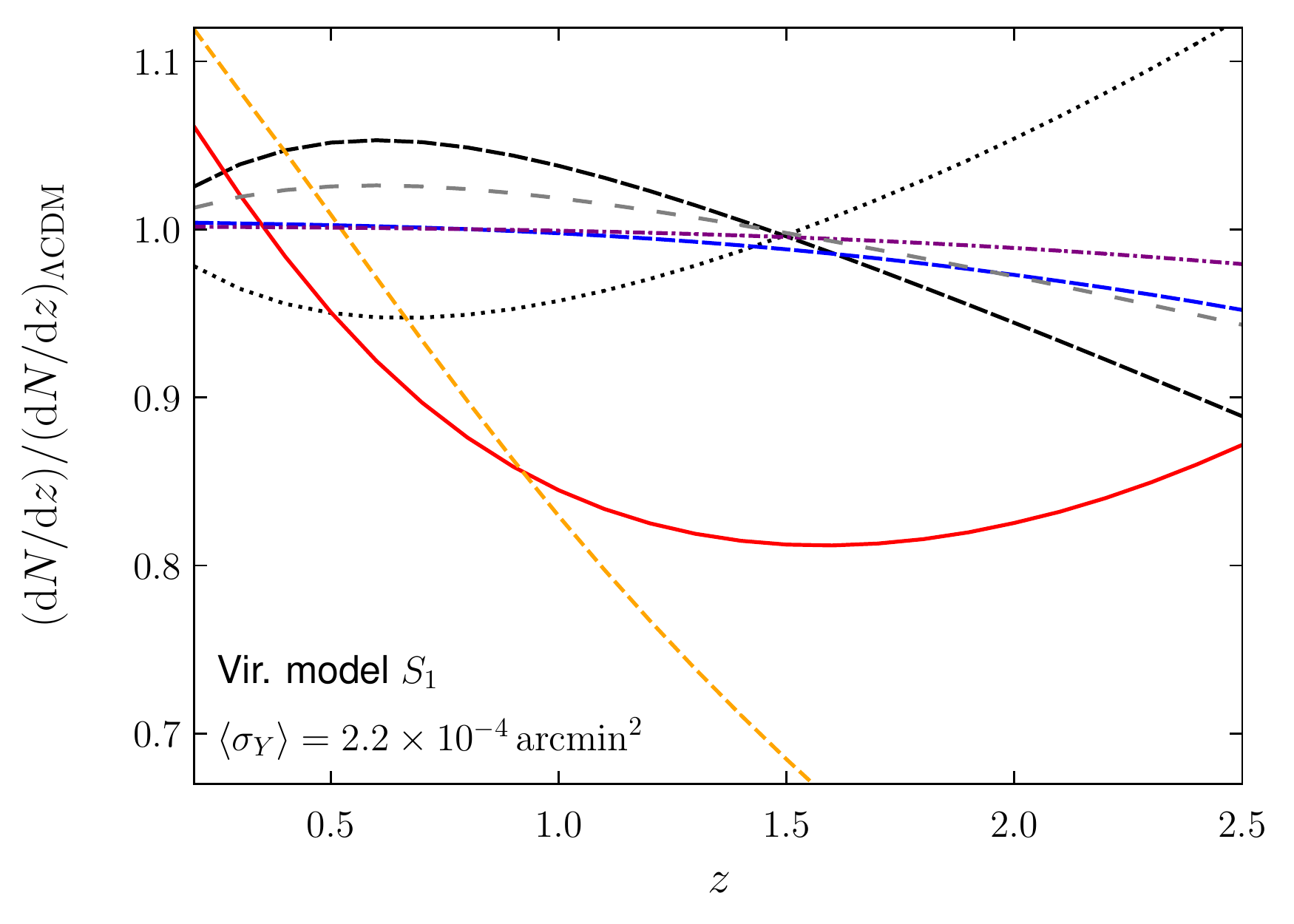}
 \includegraphics[scale=0.4]{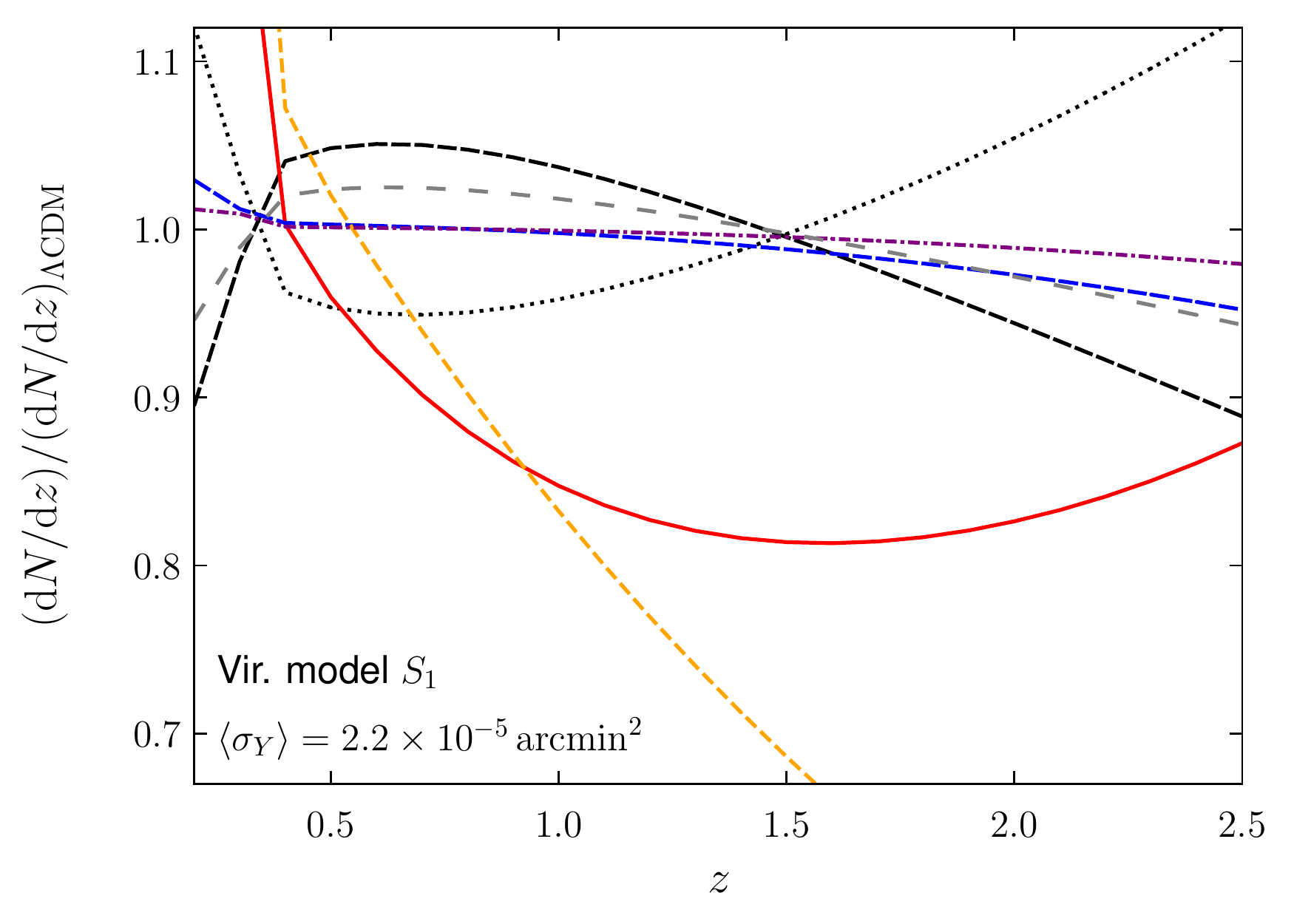}
 \includegraphics[scale=0.4]{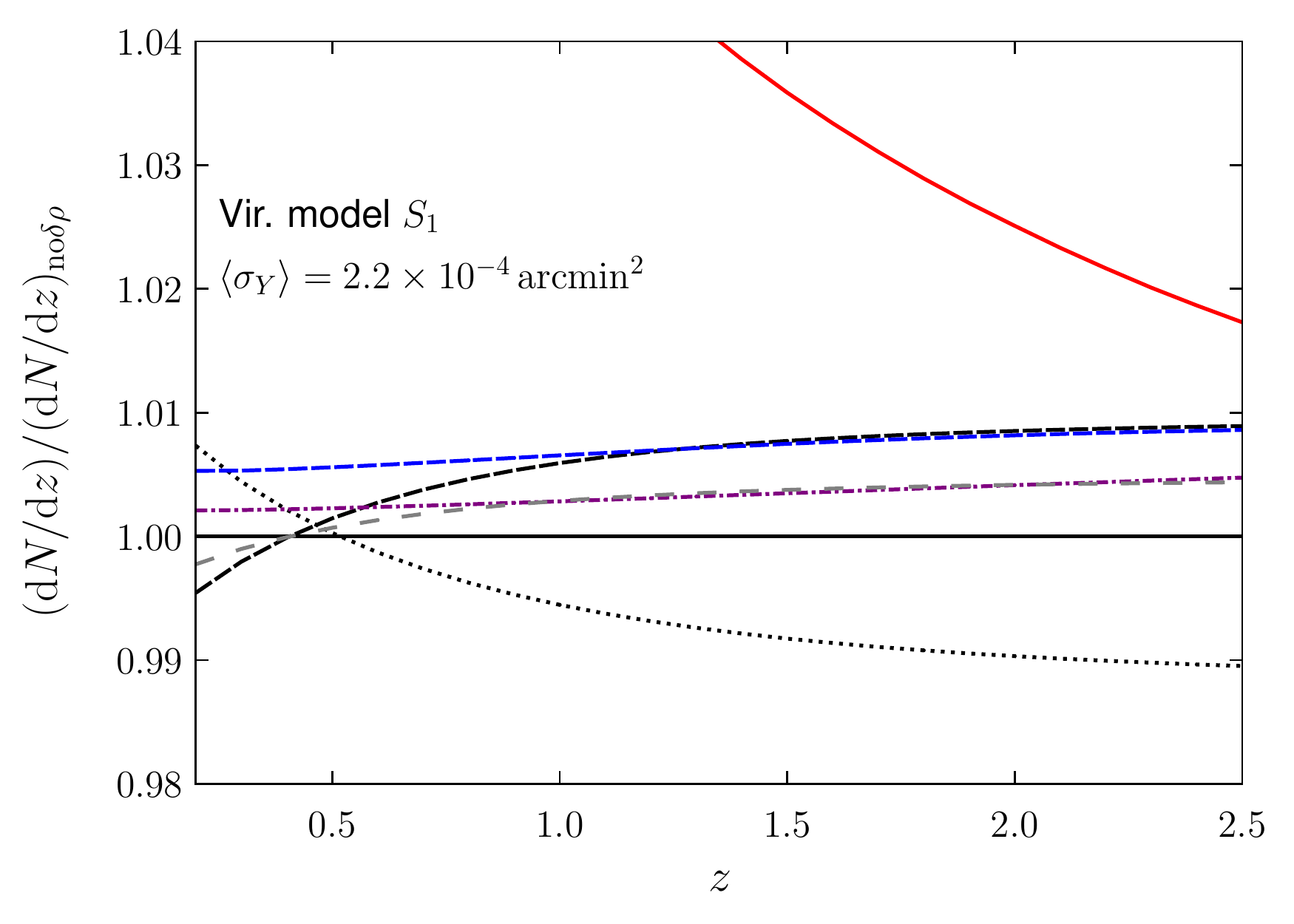}
 \includegraphics[scale=0.4]{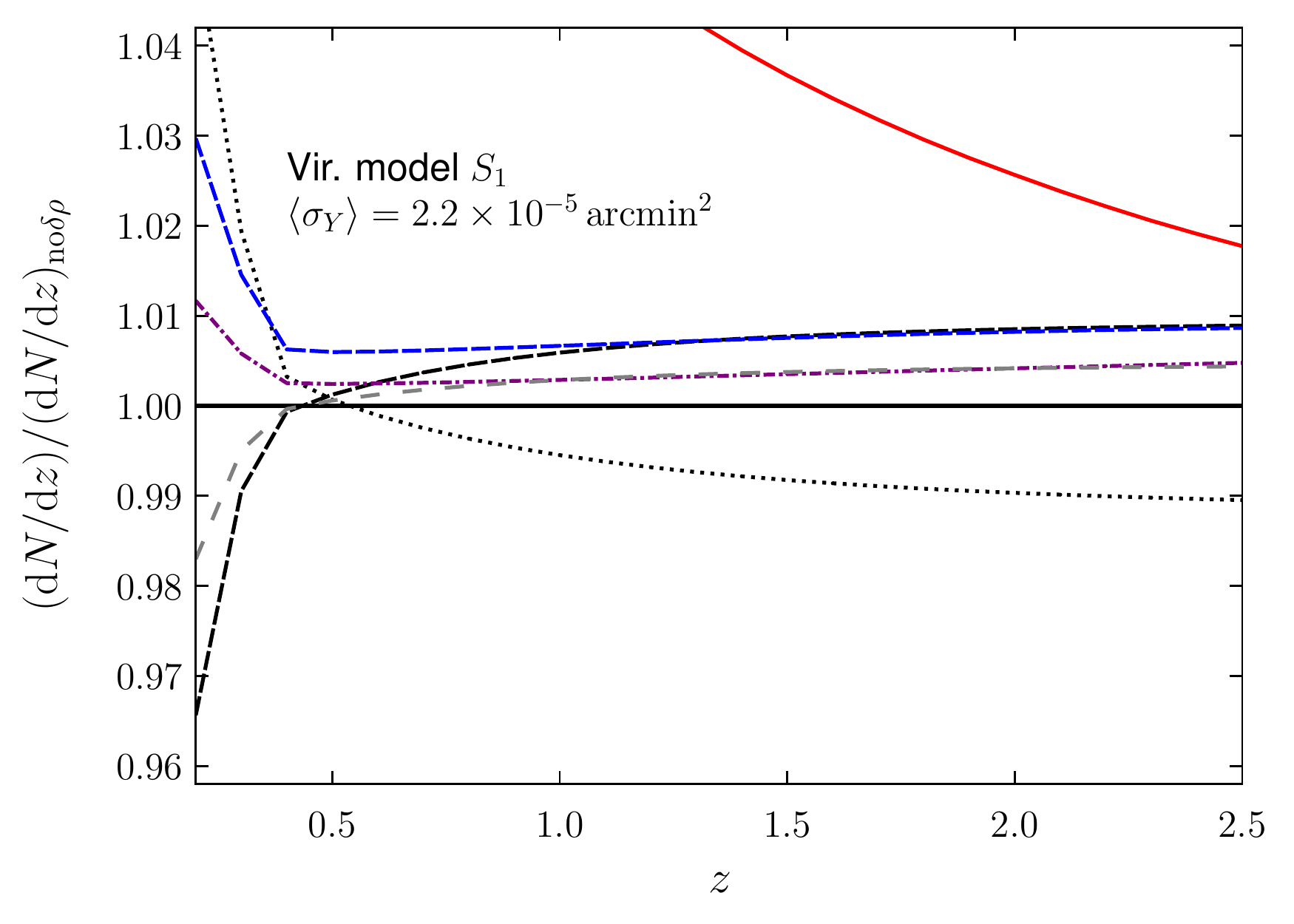}
 \caption{Differential number counts of SZ peaks in SO-like survey as function of redshift $z$. The virialization model $S_1$ and mean filter noise value $\langle\sigma_\mathrm{noise}\rangle = 2.2\times10^{-4}\mathrm{arcmin}^2$ represent the baseline (top-left panel). Counts differ by 1-4\% for the virialization model $S_2$ (top-right), by 10-15\% w.r.t. $\Lambda$CDM model (middle-left), and by less than 1\% w.r.t. smooth dark-energy models (bottom-left). All the deviations are more evident especially at low redshift with less noisy maps, if $\langle\sigma_\mathrm{noise}\rangle = 2.2\times10^{-5}\mathrm{arcmin}^2$ (middle and bottom right panels).}
 \label{fig:dNdz-SZ}
\end{figure}

\begin{table}
 \caption{Relative deviation of SZ number counts with SNR $q > 5$ and $\langle\sigma_\mathrm{noise}\rangle=2.2\times10^{-4}$arcmin$^2$ (left table) and $\langle\sigma_\mathrm{noise}\rangle=2.2\times10^{-5}$arcmin$^2$ w.r.t. $\Lambda$CDM cosmology for a SO-like CMB surveys. Quoted in parenthesis, relative deviation w.r.t. corresponding smooth dark-energy model, i.e., without perturbations ($\delta_\mathrm{de}=0$). As reference, absolute number counts with Poisson error are reported for the $\Lambda$CDM model. Values are stable against reasonable variations of cosmological parameters defining the reference model.}
 \begin{center}
 \resizebox{0.98\hsize}{!}{
  \begin{tabular}{|c|cc||cc|}
   \hline
   \cline{2-5}
    & $S_1$ & $S_2$ & $S_1$ & $S_2$ \\
   \hline\hline
    & \multicolumn{2}{c||}{$\langle\sigma_Y\rangle=2.2\times10^{-4}$arcmin$^2$} & \multicolumn{2}{c|}{$\langle\sigma_Y\rangle=2.2\times10^{-5}$arcmin$^2$}\\
   \hline
   $\Lambda$CDM & $38786\pm197$ ($0.5\%$) & $39955\pm200$ ($0.5\%$) & $68667\pm262$ ($0.4\%$) & $65234\pm255$ ($0.4\%$) \\
   \hline
   DE1 & $-2.2\%$ ($-0.5\%$) & $-2.2\%$ ($+1.4\%$) & $+4.1\%$ ($+2.0\%$) & $+3.3\%$ ($+1.8\%$) \\
   DE2 &  $+1.9\%$ ($+0.5\%$) & $+2.1\%$ ($-0.7\%$) & $-3.5\%$ ($-1.1\%$) & $-3.1\%$ ($-0.9\%$) \\
   CPL & $-13.4\%$ ($+5.3\%$) & $-13.9\%$ ($+31.7\%$) & $+18.4\%$  ($+37.3\%$) & $+16.5\%$ ($+35.4\%$) \\
   2EXP & $-0.7\%$ ($+0.7\%$) & $-0.6\%$ ($+1.7\%$) & $+0.9\%$ ($+1.7\%$) & $+0.6\%$ ($+1.6\%$) \\
   AS & $-18.8\%$ ($+13.9\%$) & $-19.4\%$ ($+57.2\%$) & $+35.4\%$ ($+72.3\%$) & $+33.2\%$ ($+69.7\%$) \\
   CNR & $-0.3\%$ ($+0.3\%$) & $-0.2\%$ ($+0.7\%$) & $+0.4\%$ ($+0.7\%$) & $+0.2\%$ ($+0.7\%$) \\
   ODE & $+0.9\%$ ($+0.3\%$) & $+1.1\%$ ($-0.3\%$) & $-1.9\%$ ($-0.6\%$) & $-1.7\%$ ($-0.5\%$) \\
  \hline
  \end{tabular}
  }
  \label{tab:SZcounts}
 \end{center}
\end{table}

Figure~\ref{fig:dNdz-SZ} shows the differential counts per unit redshift of the different clustering dark-energy models expected for a SO-like survey with Planck average noise. The top-left, middle, and bottom panels illustrate the results for the $S_1$ virialization model, the $S_2$ model yielding variations of $\sim2-3$\% monotonically increasing with redshift (top-right panel). Owing to the very large volume of such a survey and almost regardless of the value of the average noise $\langle\sigma_Y\rangle$, differential counts of clustering dark-energy would differ by $5-10\%$ w.r.t. the $\Lambda$CDM cosmology, especially at high redshift (middle panels). Dark-energy fluctuations instead only mildly affect SZ counts, which in general change by less than 1\% if $\delta_\mathrm{de}=0$ (bottom panels). The difference of counts both w.r.t. $\Lambda$CDM and smooth dark-energy models increases at low redshift ($z\lesssim0.5$) for a SZ survey with smaller noise, i.e. larger signal-to-noise ratio (middle and bottom right panels).
As for convergence peaks, all these results are similar for all but the CPL and AS models, which exhibit the strongest deviations. 

The results for integrated counts are summarised in table~\ref{tab:SZcounts}. Similarly to the table for convergence peaks counts, it reports the relative variation of SZ counts computed w.r.t. $\Lambda$CDM model and w.r.t. corresponding smooth dark-energy model (values quoted in parenthesis), for two mean noise levels of a SO-like survey. The total counts with Poissonian error (relative error in parenthesis) is reported for the $\Lambda$CDM model as reference. Also for SZ counts the depth and the sky coverage of a SO-like survey would allow us to easily distinguish the clustered dark-energy models from a $\Lambda$CDM, the only exceptions being again the 2EXP and CNR models regardless of the average noise level $\langle\sigma_Y\rangle$ and the ODE model for the smallest value of resolution. Like for convergence peaks counts, for fixed dark-energy model one can also distinguish between the two virialization models, which differ by 4 to $9\sigma$ according to the value of $\langle\sigma_Y\rangle$. Finally, like for counts of convergence peaks, the impact of dark-energy fluctuations is not negligible, especially for CPL and AS models.

\subsection{X-ray peaks: forecasts for eROSITA-like survey}
X-ray photons are mainly emitted by the core of galaxy clusters, where the local density is thousands times the critical (or mean mass) density. Here the dynamics of the dark matter is almost virialized and the memory of tidal interactions is essentially lost; we shall therefore expect very marginal dependence on the virialization model.
To integrate out any variation induced by baryonic effects very-likely occurring on such scales, one usually estimate the emission from a larger region encompassing the $R_{500}$ radius. On this scale, the exact scaling relation between the bolometric luminosity $L_X$ (in erg~s$^{-1}$), the temperature $T_X$ (in keV), and the total mass $M_{500}$ of a galaxy cluster at redshift $z$ are given by
\begin{equation}\label{eqs:LMTMscaling}
 L_X = L_\ast E^{\gamma_\mathrm{LM}}(z)\left[\frac{h}{0.72}\right]^{-0.39} \left[\frac{M_{500}}{3.9\times10^{14}M_\odot}\right]^{\alpha_\mathrm{LM}},\quad
 T_X = T_\ast \left[\frac{E(z)M_{500}}{3.02\times10^{14}h^{-1}M_\odot}\right]^{\alpha_\mathrm{TM}},
\end{equation}
with $(\log L_\ast,\gamma_\mathrm{LM},\alpha_\mathrm{LM}, T_\ast, \alpha_\mathrm{TM}) = (44.07,1.85,1.61,5,0.65)$ as typical values \cite{Vikhlinin2009a}. These relations depend on the physical properties of clusters, such as their relaxation state, metal abundance, and energy spectrum. They might also depend on $\Delta_\mathrm{vir}$, as discussed in \cite{Reichert2011} in case of self-similar evolution of haloes; we shall not consider this possibility.

The number of photons emitted by a clusters with luminosity $L_X$ and detected by an instrument in a given energy band depends on the emission spectrum of clusters, on the photoelectric absorption of X-ray photons going through the intergalactic medium, on the instrumental response of the instrument (CCD quantum efficiency, filter transmission, effective mirror area), and on the exposure time. In full generality, the number of clusters detected by more than $\eta_\mathrm{th}$ X-ray photons is given by
\begin{equation}\label{eq:Xcounts}
 N(>\eta_\mathrm{th})
 = \int \mathrm{d}\Omega \int \mathrm{d}z \int \mathrm{d}M_{500}\, \mathcal{S}(M_{500},z,\eta_\mathrm{th})\frac{\mathrm{d}V}{\mathrm{d}z\mathrm{d}\Omega}n(M_{500})\,,
\end{equation}
in which the selection function $\mathcal{S}(M,z,\eta)$ accounts for the probability of counting $\rho\equiv\eta$ photons emitted by a cluster of mass $M$ at redshift $z$. This is customarily assumed an in-homogeneous Poisson process with log-normal intensity depending on the X-ray scaling relations, namely it reads
\begin{equation}\label{eq:Xselfunction}
\mathcal{S}(M,z,\eta_\mathrm{th})\equiv P(\eta_\mathrm{th}|M,z) = \int \mathrm{d}L\,\mathrm{d}T \int_{\eta_\mathrm{th}}^\infty \mathrm{d}\eta\,P_\mathrm{Poisson}(\eta|\bar{\eta}(L,T,z))\,P(L,T|M,z)\,,
\end{equation}
where photon counts $\eta$ are a Poisson variate with mean value $\bar{\eta}(L,T,z)$ depending on the response of the instrument, on the sky scanning strategy, and on the luminosity and temperature of clusters. According to \cite{Pillepich2012,Pillepich2018}, the latter two variables are possibly correlated ($\rho_{LT}\neq0$) log-normal variates with probability $P(L,T|M,z)=P(\ln L,\ln T|\ln \bar{L}(M,z),\ln \bar{T}(M,z),\sigma_{LM},\sigma_{TM},\rho_{LT})$, with average scaling relations $\bar{L}(M,z)$ and $\bar{T}(M,z)$ given by Eqs.\eqref{eqs:LMTMscaling}.

\begin{table}
 \caption{Relative deviation of X-ray number counts w.r.t. $\Lambda$CDM for eROSITA-like all-sky survey. Quoted in parenthesis, relative deviation w.r.t. corresponding smooth dark-energy model, i.e., without perturbations ($\delta_\mathrm{de}=0$). As reference, absolute number counts with Poisson error are reported for the $\Lambda$CDM model. Values are stable against reasonable variations of cosmological parameters defining the reference model.}
 \begin{center}
  \begin{tabular}{|c|cc|}
   \hline
    & $S_1$ & $S_2$ \\
   \hline
   \hline
   $\Lambda$CDM & $144100\pm380$ ($0.3\%$) & $147810\pm384$ ($0.3\%$) \\
   \hline
   DE1 & $-3.3\%$ ($+0.2\%$) & $-3.3\%$ ($+0.2\%$) \\
   DE2 &  $+3.4\%$ ($<0.1\%$) & $+3.6\%$ ($<0.1\%$) \\
   CPL & $-4.5\%$ ($+8.1\%$) & $-4.8\%$ ($+8.2\%$) \\
   2EXP & $<0.1\%$ ($+0.5\%$) & $+0.1\%$ ($+0.5\%$) \\
   AS & $-2.7\%$ ($+15.0\%$) & $-3.2\%$ ($+15.1\%$) \\
   CNR & $<0.1\%$ ($+0.2\%$) & $+0.1\%$ ($+0.2\%$) \\
   ODE & $+1.7\%$ ($<0.1\%$) & $+1.9\%$ ($<0.1\%$) \\
  \hline
  \end{tabular}
  \label{tab:Xraycounts}
 \end{center}
\end{table}

In the following we shall consider an eROSITA-like survey operating in the 0.5–2.0 keV band, with a typical exposure time of 1.6~ks, a detection limit $\eta_\mathrm{th} = 50$ photons, and sky coverage of about 27,100~deg$^2$. Lacking for a specific model of $\bar{\eta}(L,T,z)$, we adopt the approximate model introduced in \cite{Bocquet2016} for X-ray counts based on the Magneticum simulation, which yields a redshift-dependent limiting mass $M_\mathrm{lim}(z)=\max\{4.9\times10^{13}\mathrm{M}_\odot,1.6z\times10^{14}\mathrm{M}_\odot\}$ ,i.e. a sharp cluster selection function $\mathcal{S}(M,z,\eta)=\mathcal{H}(M_{500}-M_\mathrm{lim}(z))$.

The differential counts $\mathrm{d}N/\mathrm{d}z$ as function of redshift for clustering dark-energy models resulting from this computation are qualitatively similar to SZ counts in the high-redshift tail. Interestingly enough, their relative variation w.r.t. $\Lambda$CDM and smooth dark-energy models in the redshift range $0<z<1.5$ attain almost the same values obtained for SZ over the larger range $0<z<2.5$ if $\langle\sigma_Y\rangle=2.2\times10^{-4}$arcmin$^2$. Again CPL and AS models show the largest difference, actually a factor of $\sim2$ larger than for SZ peaks' counts.

Cumulative results are summarised in table~\ref{tab:Xraycounts}. Analogously to tables~\ref{tab:SZcounts} and \ref{tab:kappapeakcounts}, absolute counts with Poisson error and relative value are reported for the $\Lambda$CDM model only and percent difference for clustering dark-energy models w.r.t. $\Lambda$CDM and smooth dark-energy models are quoted below. Also X-ray counts will allow us to clearly distinguish all but the 2EXP and CNR models, which produce counts differing by less than $1\sigma$ from numbers expected for $\Lambda$CDM. Contrary to SZ counts, taking or not into account dark-energy perturbations matters for only the CPL and AS models, and only marginally for the 2EXP model. Finally, also X-ray counts would distinguish between the $S_1$ and $S_2$ virialization models.

It is worth to note that the mass-observable relations~\eqref{eqs:LMTMscaling} of the various (clustering or smooth) dark-energy models differ by a constant bias independent of mass and depending on $E(z)$, which increases with redshift. Despite the tiny differences w.r.t. $\Lambda$CDM, they are not secondary to other physical effects which last-generation instruments are finally sensitive to. For instance, in the redshift range $0.1 \leq z \leq 0.8$ biases in cluster temperature and flux induced by the substructures of the intra-cluster medium temperature are estimated respectively of $-5.08\pm0.27\%$ and $-1.46\pm0.03\%$ for eROSITA, translating into an average mass bias of 7.5\% \cite{Hofmann2017}; at $z=0.8$ the temperature (flux) bias estimated for the dark-energy models studied here are very similar, i.e. DE1 $+1.8\%$ ($+5.1\%$), DE2 $-1.5\%$ ($-4.3\%$), CPL $+4.9\%$ ($+14.7\%$), 2EXP $<0.1\%$ ($<0.1\%$), AS $3.8\%$ ($11.2\%$), CNR $<0.1\%$ ($<0.1\%$), ODE $-0.8\%$ ($-1.2\%$). Such tiny theoretical bias could also be relevant with respect to other observational biases not considered here, e.g., induced by redshift measurement and improper classification of spectra.

\section{Conclusions}\label{sect:conclusions}

In this work, we extended the analysis of \cite{Pace2019b} by considering the effects of dark energy perturbations on virialization as induced by a shear and rotation term due to tidal forces, as originally introduced by \cite{Engineer2000} for an EdS model.

To this purpose, we investigated six dark-energy models beyond $\Lambda$CDM, dubbed DE1, DE2, CPL, 2EXP, AS, CNR, ODE (see footnote~\ref{ftn:models} for references) and extended the equations describing the time-evolution of matter perturbations taking into account dark energy perturbations, with the assumption that dark energy virialises together with matter. This seems a reasonable assumption, since density perturbations of the dominant fluid reach a constant value.

With respect to \cite{Pace2019b}, we do not discuss the evolution of the radius of perturbations nor the peculiar velocity of the shell since the qualitative behaviour does not differ from the case of smooth dark energy. We focus instead on the perturbed equation-of-state $w_\mathrm{c}$, the virial overdensity $\Delta_\mathrm{vir}$, and the effects that dark energy perturbations have on weak-lensing, SZ and X-ray counts.

The whole procedure requires to fit the amplitude of the peculiar velocity $h_{\rm SC}$ to the value measured in EdS $N$-body simulations. However, as shown in \cite{Pace2019b} the value of the free parameter $q$ is largely independent from the specific dark-energy model. For simplicity we therefore fix the value of $q$ to that found for the smooth dark-energy case and reported in table~2 of \cite{Pace2019b}. Keeping also the same background cosmology, the deviations with respect to the smooth dark-energy case are then exclusively due to dark energy perturbations.

In section~\ref{sect:epsilon} we discussed the relative contribution of dark energy to the dark matter mass component and show that the ratio $M_\mathrm{de}/M_\mathrm{m}$ is largely independent from the virialization recipe and in general is of the order of a few percent. This holds for all but the CPL and AS models, for which the dark energy contribution is up to 20\% and has a strong impact on other quantities. For the phantom models, this contribution is negative because $\delta_\mathrm{de}<0$.

This analysis allowed us to discuss the evolution of the perturbed equation-of-state $w_\mathrm{c}$ (section~\ref{sect:eos}; see figure~\ref{eqn:wc}), for which we considered two different regimes. In the first one, the system is allowed to collapse ($\delta_\mathrm{m}\to\infty$) and $w_\mathrm{c}$ goes to zero for quintessence models (as $\delta_\mathrm{de}\gg 1$), but it diverges for phantom models because $\delta_\mathrm{de}=-1$. We imposed this condition on dark energy perturbations because that value corresponds to the maximum underdensity for a void.
In the second regime, when virialization is added by suitably modifying the Euler equation, modifications to $w_\mathrm{c}$ are still present but are less strong than for complete collapse as $\delta_\mathrm{de}\simeq 1$. In this case, the effects of perturbations on the equation-of-state become important only at late times and do not depend on the virialization recipe. This is true in general for quintessence models, however for the CPL and AS models deviations from the background are still important because dark energy perturbations are relevant (see figure~\ref{fig:epsilon}). For phantom models we also observe stronger deviations from the background value, as the denominator of the expression for $w_\mathrm{c}$ becomes smaller than one.

This introductory analysis leads to the main subject of this work, namely the evolution of the virial overdensity $\Delta_\mathrm{vir}$. For this purpose, we considered two different definitions: either we define the virial overdensity as the non-linear evolution of matter perturbations only, or we explicitly include dark energy perturbations. 
In general, the two approaches give similar results, with differences at most of 2\%, except for the models CPL and AS (see figure~\ref{fig:DeltaVir}). This is due to the strong relevance of dark energy perturbations for these two models. We also considered the two different virialization recipes and showed that the one based on the perturbation radius ($S_1$) leads to smaller values of $\Delta_\mathrm{vir}$ than that based on the density perturbations ($S_2$). This is the same conclusion we obtained in \cite{Pace2019b} for smooth dark-energy models.

Non-vanishing $M_\mathrm{de}$ and new corrections to $w_\mathrm{c}$ and $\Delta_\mathrm{vir}$ induced by dark-energy fluctuations impact the number counts of density peaks as measured in gravitational lensing convergence maps, SZ maps, and X-ray surveys, collectively described by Eq.~\eqref{eq:Mcounts} that depends on the observed signal, on the mass function, and on the selection function. As benchmark, we assumed the NFW density profile and the Watson mass function \cite{Watson2013} determined by the spherical overdensity method. 
The new prescriptions modify the observables as follows:
\begin{itemize}
 \item $\Delta_\mathrm{vir}(z)$ determines the relation between the virial mass and $M_{200}$ (or $M_{500}$; see figure~\ref{fig:Mvir}), which is computed by solving the non-linear equation that depends on the density profile. As a consequence, the amplitude of the radial surface-mass-density $\Sigma(R)$ (figure~\ref{fig:Sigma1}), the (averaged) convergence $\kappa_G(\theta_G)$, Comptonization parameter $\bar{Y}_{500}$, and X-ray luminosity $L_X$ and temperature $T_X$ are altered by 1-10\% according to the dark energy and virialization models. The impact of dark energy fluctuations is marginal for all but the CPL and AS models in the range of angular scales and redshift we considered. These results are substantially unchanged if using other density profiles, such as the Einasto model.
 \item Dark-energy fluctuations enter the mass function $n(M)$ in three ways: first, via the Jacobian $\mathrm{d}M_{200}/\mathrm{d}M_\mathrm{vir}$ (or $\mathrm{d}M_{500}/\mathrm{d}M_\mathrm{vir}$); second, because we adopted the parametrization of the mass function explicitly depending on the overdensity $\Delta_\mathrm{vir}$ (see their Eqs.~17-19); third, via a multiplicative correction equal to $1+M_\mathrm{de}/M_\mathrm{vir}$ that accounts for the contribution of dark energy to the mass of the perturbations \cite{Batista2013}. For all but the CPL and AS models, the overall modification of $n(M)$ is at percent or sub-percent level with respect to $\Lambda$CDM in the mass range $10^{13.5}\,\mathrm{M}_{\odot}\,h^{-1} < M < 10^{15.6}\,\mathrm{M}_{\odot}\,h^{-1}$, with dark-energy fluctuations altering the result by $\sim1\%$ though in non-trivial way as function of mass. For CPL and AS models, the mass function differs from $\Lambda$CDM by 20-40\% for $M>10^{14}\mathrm{M}_\odot\,h^{-1}$, dark-energy fluctuations modifying the results by a factor of about 2. The modifications are independent of the virialization model.
\end{itemize}
Assuming that the simplifications for the selection functions are valid and provided one considers surveys observing very large volumes like VRO-LSST, Euclid, SO, and eROSITA, the counts of convergence, SZ, and X-ray peaks are all able to distinguish between several models of dark-energy and the $\Lambda$CDM model. Two exceptions are the 2EXP and CNR models, which presumably lead to counts very similar to those expected for the $\Lambda$CDM within Poisson error. According to the level of accuracy of our modelling, the fluctuations of dark-energy cannot be neglected. It is worth to stress that relative variations between the models are the relevant result rather than absolute values. More accurate forecasts demands a better modelling of the selection functions, tailored on specific missions and observing programs.

\acknowledgments
FP acknowledges the support from the grant ASI n.2018-23-HH.0. CS has received funding from Excellence Initiative of Aix-Marseille University - A*MIDEX, a French "Investissements d'Avenir" programme (AMX-19-IET-008 - IPhU).

\bibliographystyle{JHEP}
\bibliography{Virialization.bbl}

\label{lastpage}

\end{document}